\documentclass[twocolumn,showpacs,preprintnumbers,amsmath,amssymb]{revtex4-1}

\usepackage{graphicx} \usepackage{color} \usepackage{bm}
\usepackage{bbm} \usepackage{amsmath} \usepackage{amsfonts}
\usepackage{subfigure} \usepackage{overpic} \usepackage{float}

\newcommand{\ket}[1]{\left| #1 \right\rangle} 
\newcommand{\bra}[1]{\left\langle #1 \right|} 
\newcommand{\braket}[2]{\left\langle #1 \vphantom{#2} \right|
  \left. #2 \vphantom{#1} \right\rangle} 
 
\begin{document}
\title{Quantum walks and quantum search on graphene lattices}

\author{Iain Foulger} \author{Sven Gnutzmann} \author{Gregor Tanner}
\affiliation{School of Mathematical Sciences, University of
  Nottingham, University Park, Nottingham NG7 2RD, UK.}
  
  
\begin{abstract}

  Quantum walks have been very useful in developing search
  algorithms in quantum information, in particular for devising 
  of spatial search algorithms.  However, the construction of
  continuous-time quantum search algorithms in two-dimensional
  lattices has proved difficult, requiring additional degrees of
  freedom. Here, we demonstrate that continuous-time quantum walk
  search is possible in two-dimensions by changing the search topology
  to a graphene lattice, utilising the Dirac point in the energy
  spectrum.  This is made possible by making a change to standard
  methods of marking a particular site in the lattice. Various ways of
  marking a site are shown to result in successful search
  protocols. We further establish that the search can be adapted to
  transfer probability amplitude across the lattice between specific
  lattice sites thus establishing a line of communication between
  these sites.

\end{abstract}


\maketitle

\section{Introduction}

Quantum walks have been a particularly fruitful field of research in
quantum information going back to ideas from Feynman \cite{Feynman},
Meyer \cite{Meyer_automata}, and Aharonov \cite{Aharonov}. One focus
of recent work
on quantum walks is their application to spatial search
algorithms.  It is well-known that Grover's algorithm \cite{Grover1,
  Grover2} for searching on unstructured databases offers a quadratic
improvement in search time over classical models. Grover's algorithm
is not designed to search physical systems where only local operations
are possible and quantum walk algorithms have been employed for this
purpose.  In developing such algorithms, the major consideration is
the search time attempting to match the quadratic improvement over the
classical case offered by Grover's algorithm. The first spatial search
algorithm to do this, using the standard model of quantum walks, was
developed by Shenvi, Kempe and Whaley \cite{Shenvi} for searches on
a hypercube.

While there exist discrete-time searches on $d$-dimensional cubic
lattices which are faster than classical searches for $d\geq 2$
\cite{AKR} , effective continuous-time quantum searches only exist for
$d\geq 4$ \cite{CG04a} or else they require additional memory (in the
form of spin degrees of freedom) in order to improve their search time
in lower dimensions \cite{CG04b}. In \cite{FGT}, we demonstrated that
effective searches over two-dimensional lattices may be achieved in an
arguably simpler way which does not require extra degrees of freedom,
and could, therefore, be viewed as more efficient. This is achieved
through the choice of a different lattice, specifically, a honeycomb
lattice which is the underlying lattice structure of carbon atoms in
the material graphene.

The association with graphene is important as, although we first study
a purely theoretical problem in quantum information, the use of a
graphene lattice also offers a potential physical
realisation. 
We thus envisage using the quantum walk and quantum search algorithm
framework to investigate the effect of perturbations on the dynamics
on graphene and other carbon structures. This offers not only the
possibility for demonstrating two-dimensional continuous-time quantum
searching, but also paves the way for looking for novel effects in the
material graphene.

In this paper, we offer a detailed account of the theory and numerical
results thus expanding on the findings as presented in \cite{FGT}. 
Generalisations of the results in \cite{FGT} have been given in
\cite{Childscrystal}, where a approach to solving the problem
has been taken through encoding extra degrees of freedom into crystal
lattices. Recent experiments on artificial microwave
graphene \cite{Nice} have shown that additional site perturbations, as
discussed in Sec.\ \ref{sec:alternativemarking}, can
be used to create a search protocol. Searching on honeycomb
lattices in discrete-time setting has been considered in \cite{ADMP10}; 
however, no improvement over discrete time searches on cubic lattices 
was found.

The
paper is structured as follows: In
Section~\ref{sec:quantumwalksandsearching} we give an introduction to
the formalism of continuous-time quantum walks and the construction of
quantum walk search algorithms. We also explain why previous search
algorithms struggled on lower-dimensional lattices and why graphene
offers a solution. Section~\ref{sec:graphene} will detail the relevant
properties of graphene and set-up the notation we shall use
throughout.
Sections~\ref{sec:threebondperturbation}~\&~\ref{sec:threebondperturbation_analysis}
contain our main analytical results, where we detail the specifics of
our search algorithm and offer an analysis of the search running time
and success probability. We shall then show in
Section~\ref{sec:communication} how this search protocol can be
adapted to demonstrate novel communication
setups. Sections~\ref{sec:alternativemarking}~\&~\ref{sec:alternativenanostructures}
contain numerical work demonstrating the possibility of using
alternative methods of marking to create search behaviour and other
carbon nanostructures. We conclude with a review and discussion of our
results in Section~\ref{sec:discussion}.

\section{Quantum walks and searching}
\label{sec:quantumwalksandsearching}
Continuous-time quantum walks (CTQW), first defined in \cite{Farhi},
are the quantum analogue of continuous-time Markov chains. They are
defined purely on the state space, that is, the Hilbert space
$\mathcal{H}_{p}$, spanned by the states $\ket{j}$ which represent the
$j^{th}$ site of the lattice. Thus, the time-evolution of such systems
is defined by the Sch\"{o}dinger equation
\begin{equation}\label{ctqw}
  \frac{d}{dt}\alpha_{j}\left(t\right) = -i\sum_{l=1}^{N} {\bf H}_{jl}\alpha_{l}\left(t\right)\, ,
\end{equation}
where $\alpha_{j} = \braket{j}{\psi\left(t\right)}$ is the probability
amplitude at the $j^{\text{th}}$ vertex of a system described by the
state vector $\ket{\psi\left(t\right)}$, and $\bf H$ is the
Hamiltonian describing the connectivity of the lattice.
Note that we are using a dimensionless description setting e.g. $\hbar=1$.

The dynamics of a quantum walk over a network is defined by the nature
of the interaction between connected sites. Therefore, the Hamiltonian
is generally constructed from the adjacency matrix of the underlying
lattice. The adjacency matrix $\mathbf{A}$ is defined as
\begin{equation}
  \mathbf{A}_{jl} = \left\{ \begin{array}{l l}
      1 & \quad \text{if $j$ and $l$ are connected}\\
      0 & \quad \text{if $j$ and $l$ are not connected.}
    \end{array} \right.
\end{equation}
Typically, the Hamiltonian is chosen as $\mathbf{H} =
\epsilon_{D}\mathbf{I} + v\mathbf{A}$. The parameter $v$ determines
the coupling strength between connected sites and the parameter
$\epsilon_{D}$ is an on-site energy that only enters the dynamics in a
trivial way and thus can be set to a desired value. If $v=-1$ and $\epsilon_{D}$ is equal to the valency of the lattice $\mathbf{H}$
is the discrete Laplacian.
This form of Hamiltonian is closely related to the
tight-binding model for condensed matter systems \cite{Kittel}.

As first explained in \cite{HeinTanner09, HeinTannerCom}, a quantum
walk is transformed into a search protocol by introducing a localised
perturber state, forming an avoided crossing in the spectrum of the
search Hamiltonian between an unperturbed eigenstate and the localised
perturber. Thus, initialising the system in the unperturbed eigenstate
involved in the crossing and allowing the system to evolve in time,
one finds that the system rotates into the localised perturber state.

The first CTQW search over $d$-dimensional cubic lattices \cite{CG04a} introduced the localised perturber
state as a projector onto a single site $w$, resulting in the search
Hamiltonian
\begin{equation}\label{Hgamma}
  \mathbf{H}_{\gamma} = -\gamma \mathbf{A} + \ket{w}\bra{w}\, . 
\end{equation}
Here, $\gamma$ is a  parameter governing the strength of
interactions between sites in the lattice. This parameter is chosen carefully \cite{CG04a}  such that
the perturber state $\ket{w}$ is brought into resonance with the
ground state of the unperturbed Hamiltonian, the uniform superposition
$\ket{s}$. 

Let us assume for the moment that
all other unperturbed eigenstates are energetically
sufficiently
separated from the ground state -- we will come back to this
assumption later. Then there will be an avoided
crossing of two eigenvalues corresponding to the perturber state
and the ground state. 
Perturbation theory estimates that the energy splitting of these
resonant states is proportional to the overlap of the localised
perturber state $\ket{w}$ and the uniform ground state $\ket{s}$, which scales as
\begin{equation}
  \Delta E \sim |\langle w | s\rangle | \sim N^{-1/2}\, , 
\end{equation}
and that the corresponding eigenstates are of the form $(\ket{s} \pm \ket{w})/\sqrt{2}$.
By preparing the
system initially in $\ket{s}$ and allowing it to evolve for a set
period of time $T=\pi / \Delta E  \propto \sqrt{N}$ , a measurement of the system will result in the state
$\ket{w}$ being measured with high probability. This explains the
speed-up of the quantum walk search. The Grover algorithm and its
speed-up can be
understood in analogous terms and one can show that all other states
are indeed energetically very well separated. 

In the present case this
separation only holds above a critical dimension $d_c=4$.
A simple argument for the critical dimension can be obtained by
comparing the scaling behaviour of the energy splitting $\Delta E \sim
N^{-1/2}$ with the energy separation of the first excited state from
the ground state in the
unperturbed lattice. For a cubic lattice we have a quadratic 
dispersion relation which allows us to estimate the energy separation $$E(\underline{k}) - E_0 \sim
|\underline{k}|^2 \sim N^{-2/d}$$ (as $|\underline{k}|\sim N^{-1/d}$
for the first excited state in a $d$-dimensional lattice). For $d>4$
one then has $\Delta E/ (E(\underline{k}) - E_0) \to 0$ as $N \to
\infty$ and a detailed analysis indeed proves that the quantum walk
search works with optimal speed-up \cite{CG04a}. At the critical
dimension $d=4$ the two energy scales scale in the same way -- the
detailed analysis shows that the search still works but the dynamics
is more complicated due to the interference of excited states. 
While there is still a speed-up it is only almost optimal; the 
optimal search time $T \propto N^{1/2}$ gets multiplied with a
logarithmically
increasing factor. 
For the
experimentally relevant regimes of 2- and 3-dimensional cubic
lattices, all states participate in the dynamics as any avoided 
crossing gets dissolved completely as the number of sites grows.
As a result
no speed-up over classical searches is found.

The above estimate offers
a simple way how to reduce the critical dimension. 
If one can construct a search around a uniformly distributed state at
an energy $E_0$ where the dispersion relation is conic (linear in
$|\underline{k}|$), then in $d$ dimensions $$E-E_0 \sim
|\underline{k}| \sim N^{1/d}$$
which results in a critical dimension $d_c=2$.
In \cite{CG04b} this was implemented using a
modified Dirac Hamiltonian.  However, this requires the addition of a
spin degree of freedom, essentially a doubling of memory, with the
added complication that it is not immediately clear how such a system
would be physically realised.

Instead, our solution here is to change the lattice 
from a square (cubic) to a graphene lattice.  This change of
topology automatically implies the first step in our solution, as one
of the important electronic features of graphene is the conic
dispersion relation around the Dirac energy. This arises naturally
from the tight-binding descriptions of graphene. Note that 
graphene has the critical dimension $d=2$. Indeed our construction as 
presented in \cite{FGT} has an almost optimal speed-up with logarithmic corrections
that need to be evaluated in a detailed analysis that goes beyond the
simple perturbative description given above. 

\section{Relevant properties of graphene}
\label{sec:graphene}
\begin{figure}[t]
  \centering
  \includegraphics[width=0.45\linewidth]{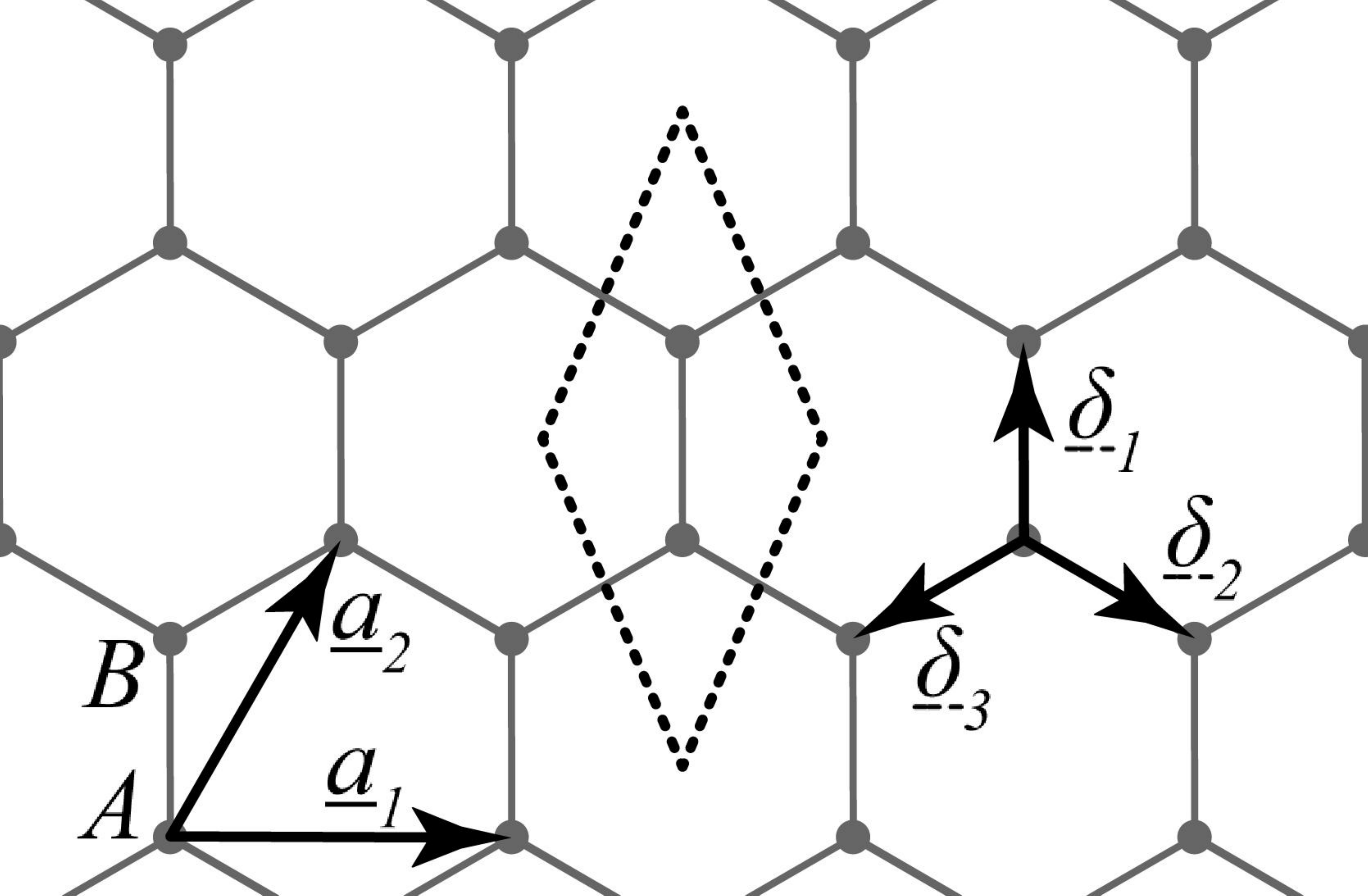}
  \hspace{0.2cm}
  \includegraphics[width=0.45\linewidth]{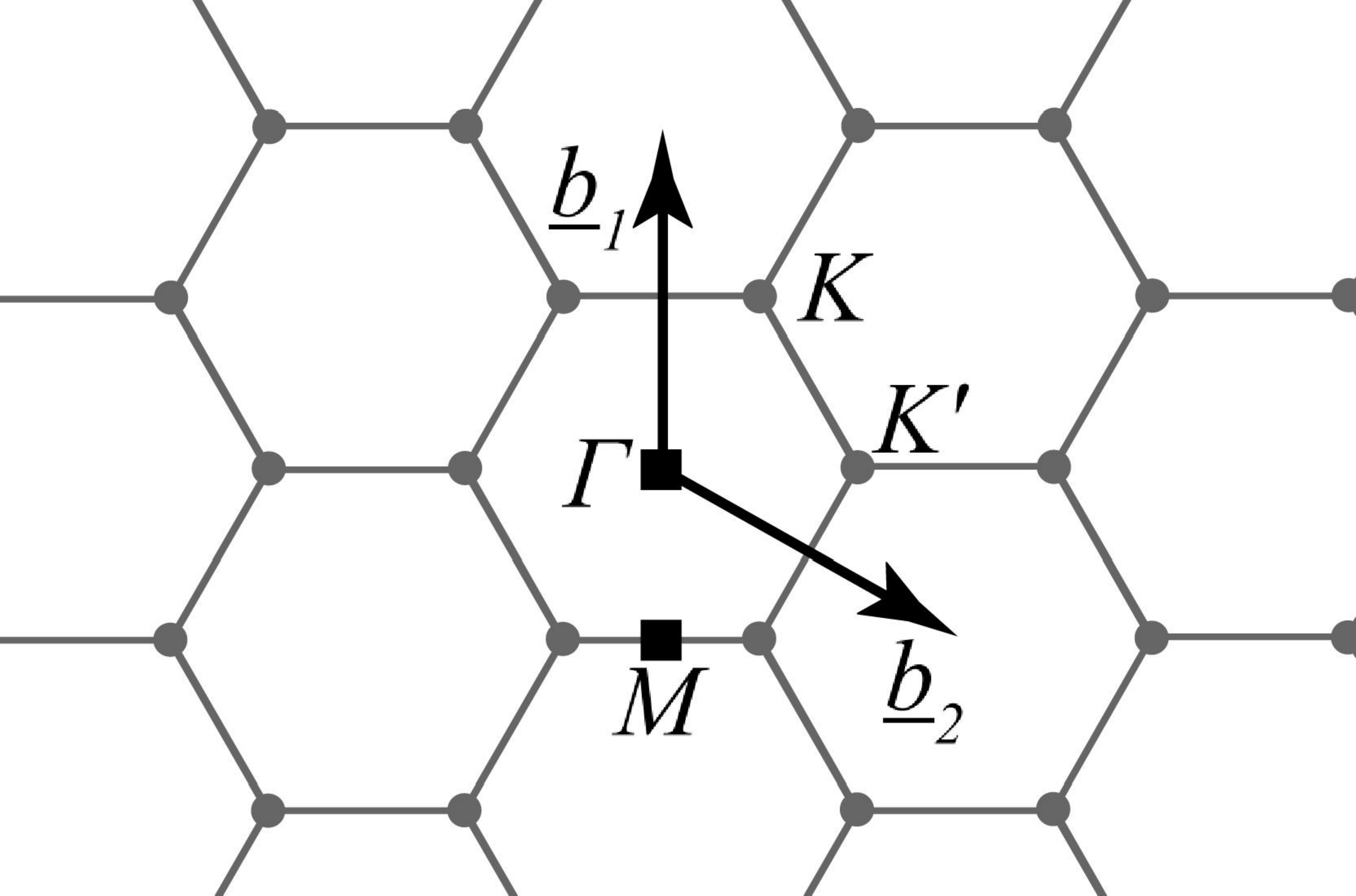}
  \label{fig:reciprocal_lattice}
  \caption{Left: Graphene with lattice vectors $\underline{a}_{1/2}$,
    translation vectors $\underline{\delta}_{i}$ and unit cell (dashed
    lines). 
    Right: Reciprocal lattice with basis vectors
    $\underline{b}_{1/2}$, symmetry points $\underline{\Gamma}$,
    $\underline{K}$, $\underline{K}'$, $\underline{M}$ and first
    Brillouin zone (hexagon).}
  \label{fig:both_lattices}
\end{figure}

Graphene is a single layer of carbon atoms arranged in a honeycomb
lattice. The lattice is bipartite with two sublattices, labelled $A$
and $B$, and a unit cell containing two carbon atoms.  The spatial and
reciprocal lattices are shown in Figure~\ref{fig:both_lattices}. The
primitive vectors describing the lattice are
$\underline{a}_{1\left(2\right)}$, such that the position of a unit
cell in the lattice is given by
$\underline{R}\left(\alpha,\beta\right) = \alpha a_{1} + \beta
a_{2}$. We use dimensionless units in space where the
distance between nearest neighbor sites is $a=1$.
The reciprocal lattice shows two important points, the Dirac
points $\underline{K}$ and $\underline{K}'$ at the two inequivalent
corners of the Brillouin zone.

\begin{figure}
  \centering
  \includegraphics[trim = 0mm 63mm 0mm 92mm,clip = true,
  width=0.9\linewidth]{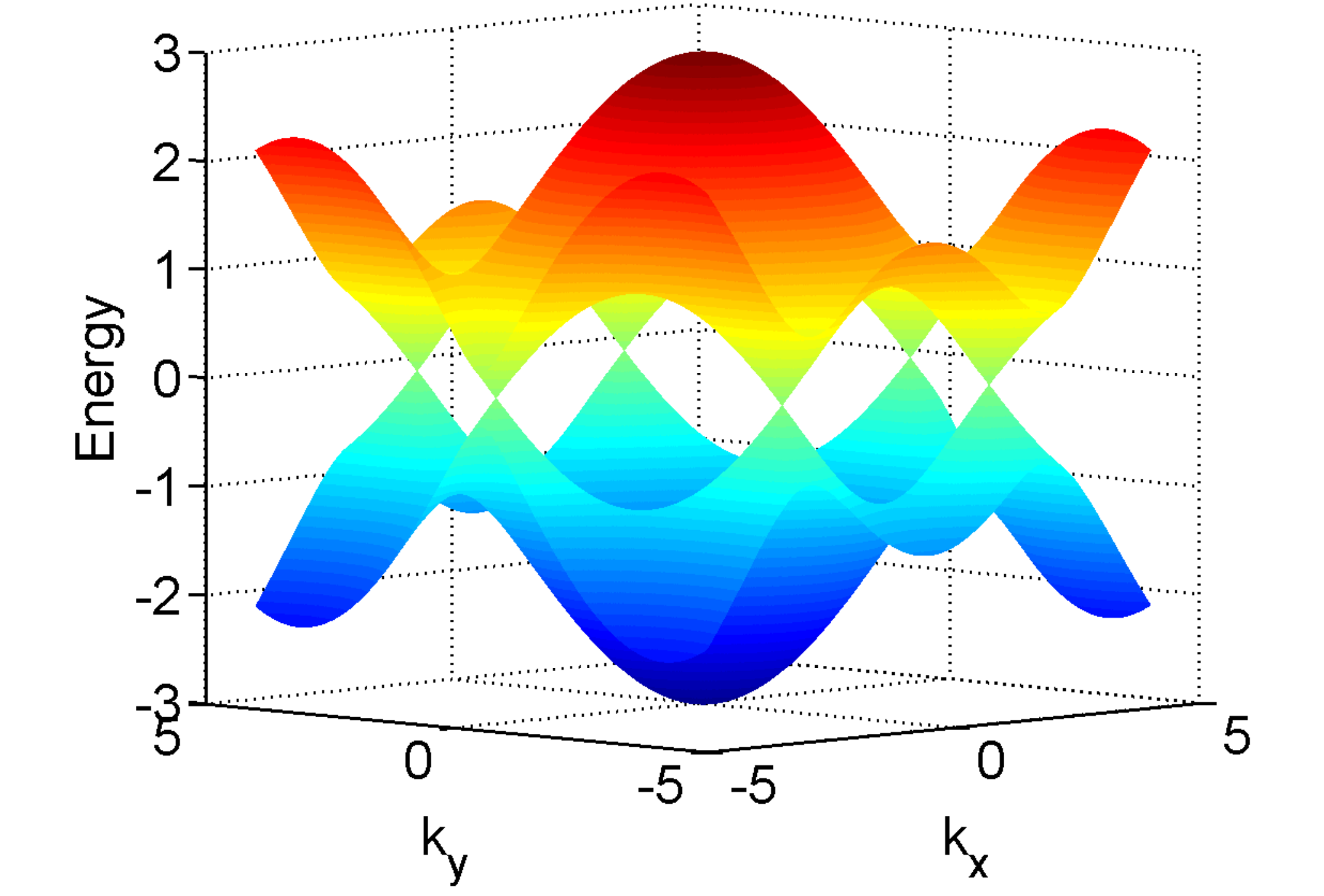}
  \caption{(Color online) Dispersion relation for infinite graphene
    sheet ($\epsilon_{D}=0$ and $v=-1$).}
  \label{fig:infinite_graphene_lattice}
\end{figure}

The energy spectrum of electrons in graphene was first derived by
Wallace \cite{Wallace_graphene} when considering the band structure of
graphite using a tight-binding Hamiltonian
\begin{equation}
  \mathbf{H}= \epsilon_D\mathbf{1} + v \mathbf{A}
\end{equation}
where $\epsilon_D$ is the on-site energy (which we will identify as the energy
of the Dirac points) and
$v$ the hopping strength (both dimensionless in our setting).
The tight-binding model
for graphene and the derivation of the solution are well-known
\cite{Wong_nanotube, Neto_graphene} and give rise to the dispersion
relation
\begin{align}
  \epsilon\left(\underline{k}\right) &= \epsilon_{D} \pm \label{eq:graphene_unperturbed_eigenenergies} \\
  &v\sqrt{1 + 4\cos^{2}\left(\frac{k_{x}}{2}\right) +
    4\cos\left(\frac{k_{x}}{2}\right)\cos\left(\frac{\sqrt{3}k_{y}}{2}\right)}\,
  , \nonumber
\end{align}
shown in Figure~\ref{fig:infinite_graphene_lattice} for an infinite
graphene lattice. 

As there are two atoms per unit cell the spectrum has two branches,
the upper branch being the conduction band and the lower the valence
band, which meet at the corners of the Brillouin zone, the
$\underline{K}$-points. The energy at the $\underline{K}$-points is
$\epsilon_{D}$ which we name the Dirac energy. It is around these
points, that the behaviour of the spectrum is conical, that is,
\begin{equation}\epsilon\left(\underline{k}\right) \approx
  \epsilon_{D} \pm v\frac{\sqrt{3}}{2}\sqrt{\delta k_{x}^{2} + \delta
    k_{y}^{2}} = \epsilon_{D} \pm v\frac{\sqrt{3}}{2}\left|\delta
    k\right|\, ,
  \label{eq:linear_dispersion}
\end{equation}
with a reduced density of states, a necessary feature for the creation
of the search dynamics.

As the lattice possesses a translational symmetry the Hamiltonian can
be solved using linear superpositions of Bloch functions over both
sublattices. As a basis we use the orthonormal states
$\{\ket{\alpha,\beta}^{A},\ket{\alpha,\beta}^{B}\}$ to denote states
on either the $A$ or $B$ sublattice in the cell at position
$\underline{R}\left(\alpha,\beta\right)$.  For the majority of what
follows (except in Section~\ref{sec:alternativenanostructures}), we
will focus on finite-sized lattices with assumed periodic boundary
conditions along the axes of both primitive vectors so that the
topology of our lattice is a torus; that is, our wavefunction is of
the general form $|\psi\rangle= \sum_{\alpha=1}^m\sum_{\beta=1}^n
\left(\psi_{\alpha,\beta}^A |\alpha,\beta\rangle^A + \allowbreak
  \psi_{\alpha,\beta}^B |\alpha,\beta\rangle^B\right)$. Our boundary
conditions imply that the state vector must satisfy
$\psi^{A(B)}_{\alpha,\beta}=\psi^{A(B)}_{\alpha+m,\beta}=\psi^{A(B)}_{\alpha,\beta+n}$
where $m,n$ denote the period of the lattice. Thus, the wavefunctions
on the torus take the Bloch function form
\begin{align}
  \ket{\psi} = &\sum_{\left(\alpha,\beta\right)} \left[ \frac{1}{\sqrt{N}}e^{i\underline{k}\cdot\underline{R}\left(\alpha,\beta\right)}\ket{\alpha,\beta}^{A} \right. \nonumber \\
  &\left. + \frac{C\left(\underline{k}\right)}{\sqrt{N}}
    e^{i\underline{k}\cdot\left(\underline{R}\left(\alpha,\beta\right)+\underline{\delta}_{1}\right)}\ket{\alpha,\beta}^{B}\right]
  \, , \label{eq:wavefn}
\end{align}
where $N$ is the number of sites in the lattice, $\underline{k}$ is
the momentum, and $C\left(\underline{k}\right)$ is a relative phase
contribution dependent upon whether the state belongs to the
conduction or valence bands; it can be calculated by explicitly
working through the tight-binding model.

Application of the periodic boundary conditions results in the
following quantised momenta
\begin{equation}
  k_{x} = \frac{2\pi p}{m}\, , \hspace{10mm} k_{y} = \frac{1}{\sqrt{3}}\left(\frac{4\pi q}{n} - k_{x}\right)\, ,
  \label{eq:quantised_momenta}
\end{equation}
where $p \in \{0,1,\ldots m-1\}$ and $q \in \{0,1,\ldots n-1\}$.  In
the following and whenever we consider quantum walk dynamics on a
torus, the number of cells in each direction is generally chosen to be
the same, that is, $m = n = \sqrt{\frac{N}{2}}$.  This choice is
purely for simplifying the notation; alternative torus dimensions are
possible as are other choices of boundary conditions
corresponding to alternative carbon structures (e.g.\ nanotubes or a
graphene sheet) as will be demonstrated in
Section~\ref{sec:alternativenanostructures}. For our choice of torus
dimensions, we find there are momenta equal to the
$\underline{K}$-points and, consequently, eigenstates with energies
equal to the Dirac energy when both $m$ and $n$ are some multiple of
3.

In fact, using the quantised momenta in
Eq.~\eqref{eq:quantised_momenta} obtained for periodic boundary
conditions, we find that there are four degenerate eigenstates with an
energy that coincides exactly with the Dirac energy when $m$ and $n$
are both multiples of 3. These four states, known as the Dirac states,
can be constructed to live only on one of the sublattices, and are given by
\begin{align}\label{eq:K_state}
  \ket{\underline{K}}^{A(B)} =& \sqrt{\frac{2}{N}} \sum_{\left(\alpha,
      \beta\right)} e^{i\frac{2\pi}{3}\left(\alpha + 2\beta + 2\sigma
    \right)}
  \ket{\alpha,\beta}^{A(B)}\nonumber \\
  \ket{\underline{K}'}^{A(B)} =&\sqrt{\frac{2}{N}} \sum_{\left(\alpha,
      \beta\right)} e^{i\frac{2\pi}{3}\left(2\alpha +
      \beta\right)}\ket{\alpha,\beta}^{A(B)}\, ,
\end{align}
where $\sigma = 0$ for states on the $A$-sublattice or $\sigma = 1$
for states on the $B$-sublattice.

\section{Quantum search on graphene - Reduced model}
\label{sec:threebondperturbation}
In this section we will describe how a site is marked and will derive
our optimal search starting state. Our approach will be to analyse the
system's spectrum and its dynamics in a reduced Hamiltonian model
involving only the relevant states from the avoided crossing. In the
next section we will then validate this approach with a more detailed
analysis, using the results obtained here as an initial guide.

As already established, we introduce the localised perturber state in
a region of the spectrum with a conic dispersion relation and thus a
low density of states. Simply altering the on-site energy in Eq.\
(\ref{Hgamma}) as done in \cite{CG04a} does, however, not work here.
As the on-site energy $\epsilon_{D}$ and the Dirac energy are equal,
one finds that the perturbation only interacts with the Dirac states
in the limit of zero perturbation strength, returning the unperturbed
lattice.  Therefore, we choose an alternative perturbation method:
namely, we modify the coupling strength between the site we wish to
mark and its neighboring vertices.  We focus in this section on
changing the coupling to all three neighboring vertices of a
particular site equally. Our choice of perturbation matrix $\bf{W}$ to
mark the $A$-type vertex $\left(\alpha_{o},\beta_{o}\right)^{A}$ is
then
\begin{equation}\label{eq:3bond_perturbation_matrix} {\bf{W}} =
  \sqrt{3}\ket{\ell}\bra{\alpha_{o},\beta_{o}}^{A} +
  \sqrt{3}\ket{\alpha_{o},\beta_{o}}^{A}\bra{\ell}\, ,
\end{equation}
where the state $\ket{\ell}$ is the symmetric superposition over the
three neighbors of the perturbed site
\begin{equation}
  \ket{\ell} = \frac{1}{\sqrt{3}}\left(\ket{\alpha_{o}-1,\beta_{o}}^{B} + \ket{\alpha_{o}-1,\beta_{o}+1}^{B} + \ket{\alpha_{o},\beta_{o}}^{B} \right)\, .
\end{equation}
This leaves us with the search Hamiltonian
\begin{equation}
  \bf{H}_{\gamma} = -\gamma \bf{A} +  \bf{W}\, ,
  \label{eq:full_search_Hamiltonian}
\end{equation}
where $\gamma$ is a free parameter. In what follows, we always set the
on-site energy  $\epsilon_{D}=0$.
Considering our search Hamiltonian, we can see that setting $\gamma =
1$ corresponds to a coupling strength of $v=0$ from the perturbed
vertex and its nearest-neighbors; our perturbation essentially
removes the marked vertex $\left(\alpha_{o},\beta_{o}\right)^{A}$ from
the lattice.  Note that vacancies are a common, naturally occurring,
defects in graphene lattices \cite{Meyer}.

\begin{figure}
  \begin{overpic}[width =
    1.05\linewidth]{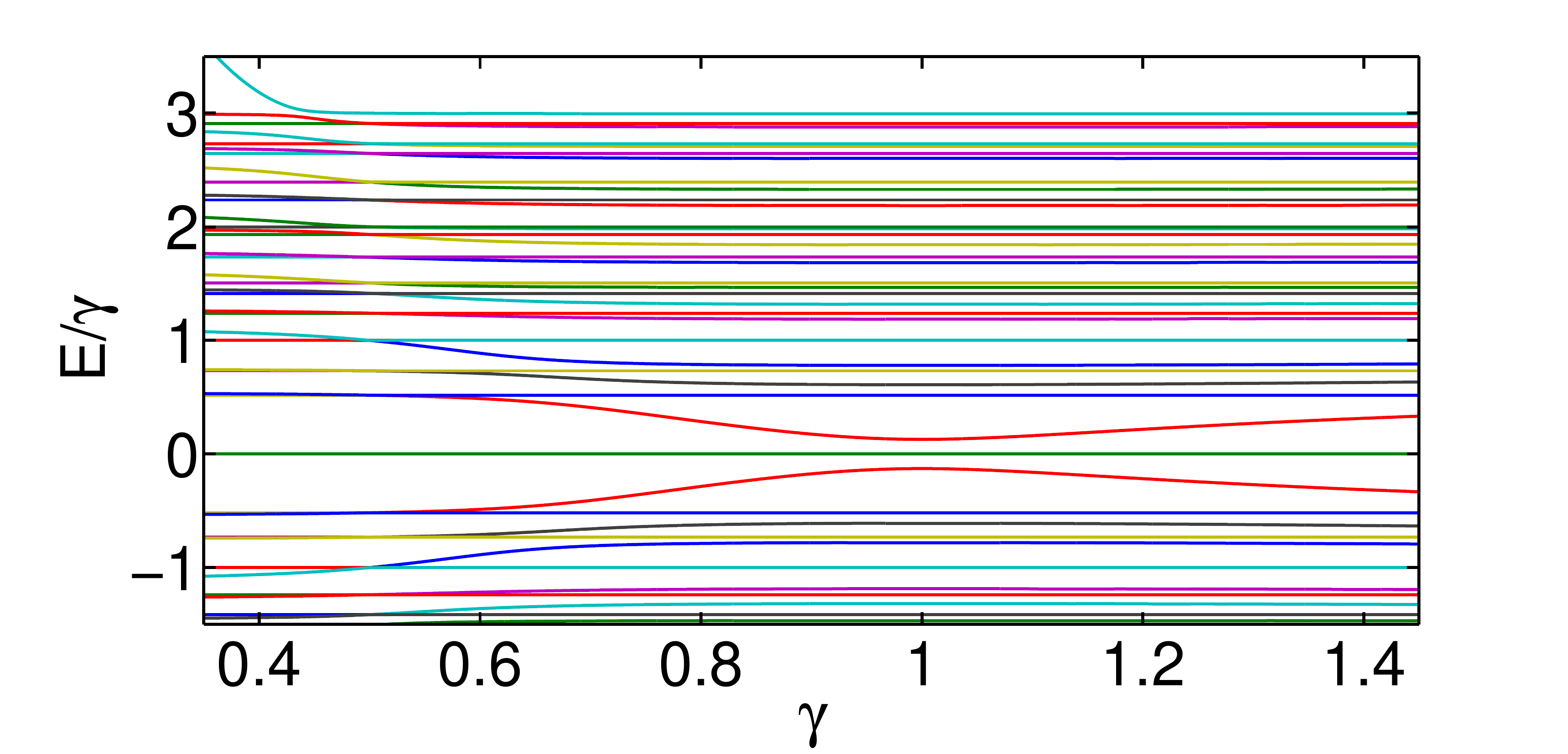}
    \put(54.25,26.75){\fbox{\includegraphics[width=0.37\linewidth]{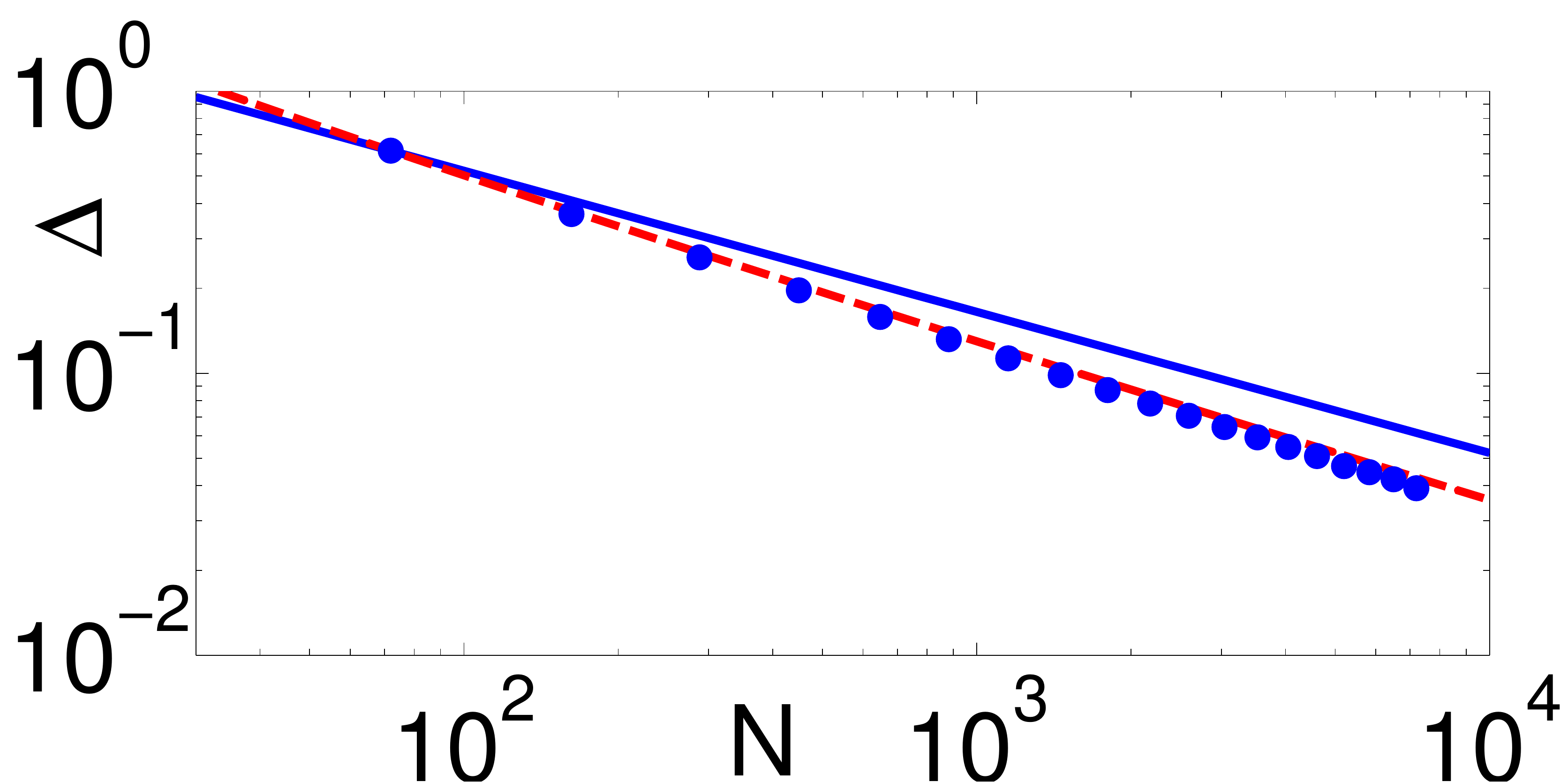}}}
  \end{overpic}
  \caption{(Color online) Spectrum of $ {\bf H}_\gamma$ in
    Equation~\eqref{eq:full_search_Hamiltonian} as a function of
    $\gamma$ for a $12\times 12$ cell torus ($N=288$). The spectrum is
    symmetric around $\epsilon_{D} = 0$.  Inset: Scaling of the gap
    $\Delta = \tilde{E}_{+} - \tilde{E}_{-}$ (dots) and curves
    $c_{1}/\sqrt{N}$ (solid blue), $c_{2}/\sqrt{N\log N}$ (dashed red)
    for comparison. }
  \label{fig:3bond_spectrum}
\end{figure}

In order to establish the critical value of $\gamma$, we numerically
calculate the spectrum of $\bf{H}_{\gamma}$ as a function of $\gamma$,
plotted in Figure~\ref{fig:3bond_spectrum} for a torus of dimensions
$m = n = 12$.  As $\bf{W}$ is a rank-2 perturbation, we see in
Figure~\ref{fig:3bond_spectrum} two perturber states working their way
through the spectrum to an avoided crossing around $\epsilon_{D} = 0$
when $\gamma = 1$, that is, when the perturbed vertex is removed from
the underlying lattice.

As well as establishing a critical value for $\gamma$, we also find
from this figure the states involved in the avoided crossing: the four
Dirac states at the Dirac energy and the two perturber states which
form our perturbation matrix $\bf{W}$.  However, with further
consideration, we may reduce the number of states involved further. As
we have removed the site $\left(\alpha_{0},\beta_{0}\right)^{A}$ from
the lattice it can no longer interact with the rest of the lattice and
the corresponding state drops out. Also, by direct calculation one
finds that the $B$-type Dirac states do not interact with the
perturbation, that is, ${\bf{W}}\ket{\underline{K}}^{B} =
{\bf{W}}\ket{\underline{K}'}^{B} = 0$. Thus, they remain an eigenstate
of the search Hamiltonian and do not interact with the perturbation.
We are left with three states taking part in the avoided crossing:
$\{\ket{\underline{K}}^{A},\ket{\underline{K}'}^{A},\ket{\ell}\}$.

We reduce our search Hamiltonian in this three state basis at the
critical point $\gamma = 1$ to obtain the following reduced
Hamiltonian describing the local dynamics at the avoided crossing
\begin{equation}\label{redH}
  \tilde{\bf H} = \sqrt{\frac{6}{N}} \begin{bmatrix}
    0 & 0 & e^{-i\mu_o}  \\
    0 & 0 & e^{-i\nu_o} \\
    e^{i\mu_o} & e^{i\nu_o} & 0   
  \end{bmatrix}\,
\end{equation}
with $\mu_o = \frac{2\pi}{3}\left(\alpha_{o} + 2\beta_{o}\right)$ and
$\nu_o = \frac{2\pi}{3}\left(2\alpha_{o} + \beta_{o}\right)$.  This
reduced Hamiltonian has eigenvalues $ \tilde{E}_{\pm} = \pm
2\sqrt{\frac{3}{N}}, \tilde{E}_{0} = 0$, and eigenvectors
\begin{align} \label{eq:reduced_matrix_eigenvectors}
  \ket{\tilde{\psi}_{\pm}} &= \frac{1}{2}\left(e^{-i\mu_o}\ket{K}^A + e^{-i\nu_o}\ket{K'}^A \pm \sqrt{2}\ket{\ell}\right) \\
  \ket{\tilde{\psi}_{0}} &=
  \frac{1}{\sqrt{2}}\left(e^{-i\mu_o}\ket{K}^A - e^{-i\nu_o}\ket{K'}^A
  \right)\, .
\end{align}

Using the eigenvectors of the reduced Hamiltonian, we can construct a
search starting state which is a superposition of Dirac states

\begin{align}
  \ket{s} &= \frac{1}{\sqrt{2}}\left(\ket{\tilde{\psi}_{+}} + \ket{\tilde{\psi}_{-}}\right) \label{eq:starting_state} \\
  &= \frac{e^{-i\mu_o}}{\sqrt{2}}\left(\ket{\underline{K}}^{A} +
    e^{-i\frac{2\pi}{3}\left(\alpha_{o} -
        \beta_{o}\right)}\ket{\underline{K}'}^{A}\right)\, . \nonumber
\end{align}

Allowing our search starting state $\ket{s}$ to evolve under the
reduced Hamiltonian we find

\begin{align}
  \ket{\psi\left(t\right)} &= e^{-i{\bf{\tilde{H}}}t}\ket{s} \nonumber \\
  &= \frac{1}{\sqrt{2}}\left(e^{-i \tilde{E}_{+}t}\ket{\tilde{\psi}_{+}} + e^{-i \tilde{E}_{-}t}\ket{\tilde{\psi}_{-}}\right) \label{eq:search_time}\\
  &= \cos\left(\tilde{E}_{+}t\right)\ket{s} -
  i\sin\left(\tilde{E}_{+}t\right)\ket{\ell} \nonumber\, ,
\end{align}
so that our system rotates from our Dirac superposition $\ket{s}$ to a
state which is localised on the neighbors of the perturbed vertex
$\ket{\ell}$ in a time $t=\frac{\pi}{4}\sqrt{\frac{N}{3}}$. Thus, we
find a $\mathcal{O}\left(\sqrt{N}\right)$ search time, a polynomial
improvement over the classical search time.

\begin{figure}[t]
  \centering
  \includegraphics[width= 1.0\linewidth]{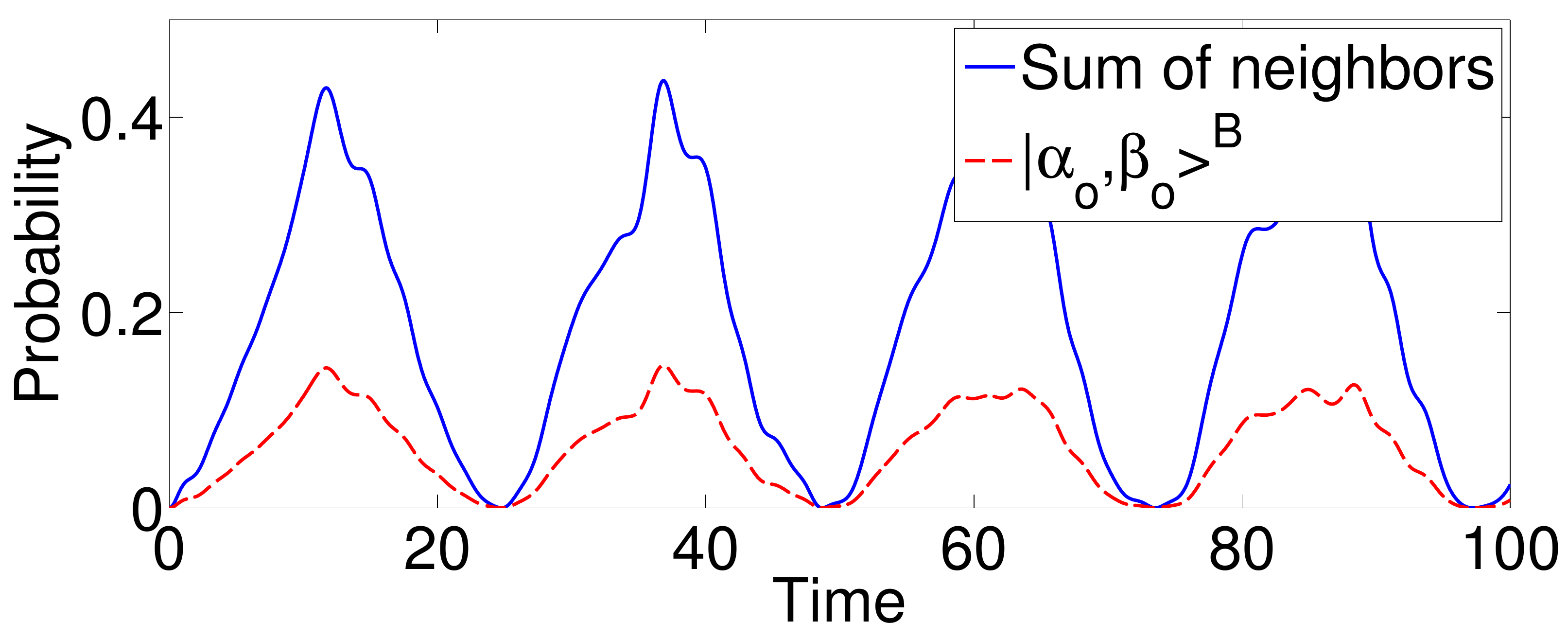}
  \caption{(Color online) Search on $12\times 12$ cell graphene lattice with a
    triple-bond perturbation, using starting state $\ket{s}$ from
    Equation~\ref{eq:starting_state}. For tori with $m = n$ the
    dynamics at each neighboring site is the same so only one is
    shown.}
  \label{fig:search_prob}
\end{figure}

We plot in Figure~\ref{fig:search_prob} a numerically calculated
search for a lattice with $N = 288$ sites. The system has been
prepared in $\ket{s}$ and allowed to evolve under the full search
Hamiltonian from Eq.~(\ref{eq:full_search_Hamiltonian}) at the
critical value $\gamma = 1$.  One finds that the system localises on
the neighbors of the marked vertex and, for this particular system
size, the probability for a measurement of the system to return one of
the neighbors is around $45\%$ at its peak.  This is orders of
magnitude higher than the average probability to measure other
vertices in the lattice, $100/N$, which for this system size is around
$0.35\%$.  The localised state interacting with the spectrum under the
full Hamiltonian has a tail into the rest of the lattice; this leads
to a loss of probability to be found at the neighboring vertices
below $100\%$.

It is worth emphasising that the marked vertex plays no role in the
search - it is removed from the lattice. Rather, we force the system
to localise on the nearest-neighbors, essentially making the marked
vertex conspicuous by its absence.  This feature differentiates our
algorithm from previous searches, which have generally concentrated on
localising directly on the marked vertex. An approach considering the
neighboring vertices was detailed in \cite{Ambainisneighbors},
however, the focus was here still on localising on a single site but
with classical post-processing steps considering the tail of the
localised state extending into the neighborhood around the marked
vertex.

At this point, there are a number of issues which need to be
mentioned. The first is that our starting state in
Eq.~(\ref{eq:starting_state}) contains information about the marked
vertex in the form of the relative phase between the Dirac states.
However, this phase can only take three different values.  Therefore, we have
three possible optimal starting states for an $A$-type perturbation and the
same number for $B$-type perturbations; there are thus in total six
possible optimal states.  As not all of these states are orthogonal,
we only find an increase of necessary runs for a successful search by
a factor of 4; this additional overhead is independent of $N$ and,
therefore, does not affect the overall time complexity of the search.
The particular representation of the Dirac states in
Eq.~(\ref{eq:K_state}) was chosen in such a way as to make the
calculations and conceptualisation of the system dynamics easier. In
reality, constructing starting states  which exist on
one sublattice only, experimentally is extremely difficulty; rather, in an experiment,
one is likely to excite a superposition of Dirac states, reducing the
success probability, on average, by a factor of $1/4$.

Our search is based on the conic dispersion relation in the spectrum
and the $\mathcal{O}\left(1/\sqrt{N}\right)$ scaling between
successive energy levels in the linear regime, giving rise to the
$\mathcal{O}\left(\sqrt{N}\right)$ search time found in our reduced
model.  However, previous searches in continuous-time using a modified
Dirac Hamiltonian \cite{CG04b} or in discrete-time \cite{AKR} have
found logarithmic corrections to the search time.  As we have seen in
Eq.~(\ref{eq:search_time}) the search time is related to the energy
gap at the avoided crossing. In the inset of
Figure~\ref{fig:3bond_spectrum} we have numerically calculated the
scaling of the gap in the spectrum of the full search Hamiltonian and
we indeed find evidence for a logarithmic correction. The lack of a
logarithmic term in the search time for our reduced model is due to
the neglecting contributions from the rest of the spectrum.

In what follows we establish a more accurate estimate of the running
time and success probability by working with the full Hamiltonian. We
also justify the findings of our reduced model: while this model is
insufficient to estimate the finer details, such as logarithmic correction terms, 
the accurate calculations show that our search
algorithm indeed takes place mainly in a two-dimensional subspace
spanned by the Dirac states and a state localised on the neighbors of
the marked vertex. 

\section{Quantum search on graphene - Detailed analysis}
\label{sec:threebondperturbation_analysis}

We follow here closely our derivation as  presented in \cite{FGT} which builds on ideas given in \cite{CG04a} for a similar 
a similar calculation for regular rectangular lattices. We focus first on the search success amplitude, that is,
\begin{equation}
  \bra{\ell}e^{-i{\bf{H}}T}\ket{start} = \sum_{\ket{\psi_{a}} 
  } \braket{\ell}{\psi_{a}}\braket{\psi_a}{start}e^{-iE_{a}T},
  \label{eq:amplitude}
\end{equation}
where $T$ is the search time, that is, the time our search probability
reaches a maximum, and our search Hamiltonian ${\bf H}$ is given by
Eq.\ (\ref {eq:full_search_Hamiltonian}) with $\gamma = 1$, that is,%
\begin{equation} {\bf{H}} = -{\bf{A}} +
  \sqrt{3}\ket{\ell}\bra{\alpha_{o},\beta_{o}} +
  \sqrt{3}\ket{\alpha_{o},\beta_{o}}\bra{\ell}\, ;
  \label{eq:specific_search_Hamiltonian}
\end{equation}
here, $\ket{\psi_a}$, $E_{a}$ are the eigenstates and eigenenergies of
${\bf H}$.  We assume, without loss of generality, a specific
starting state $\ket{start}$ where the marked vertex is chosen such
that $e^{i\frac{2\pi}{3}\left(\alpha_{o}+2\beta_{o}\right)} = 1$
leading to
\begin{equation}
  \ket{start} = \frac{1}{\sqrt{2}}\left(\ket{\underline{K}} + \ket{\underline{K}'}\right)\, .
\end{equation}
In the following, we suppress the sublattice superscript as the
analysis is the same regardless on which sublattice the perturbation
lives. We also denote $\epsilon\left(\underline{k}\right)$ the positive 
eigenenergies of $-\bf{A}$ from Eq.~(\ref{eq:graphene_unperturbed_eigenenergies}) 
and set $\epsilon_{D} = 0$. 

For eigenstates $\ket{\psi_{a}}$ with eigenenergies $E_a$
that are not in the unperturbed graphene spectrum ($E_a \neq
\epsilon(\underline{k})$ for all points $\underline{k}$ in the dual lattice),
we may rewrite
Eq.~(\ref{eq:specific_search_Hamiltonian}) in the form
\begin{equation}
  \ket{\psi_{a}} = \sqrt{3 R_{a}}(E_{a} + {\bf{A}})^{-1} \ket{\alpha_{o},\beta_{o}}\, ,
  \label{eq:eigenstate_representation}
\end{equation}
where $\sqrt{R_{a}} = \braket{\ell}{\psi_{a}}$ and the phase of
$\ket{\psi_{a}}$ is chosen such that $\braket{\ell}{\psi_{a}} \geq 0$.

At this point, we can remove several states from the summation in
Eq.~(\ref{eq:amplitude}).  Note that the basis state associated with
the marked vertex, $\ket{\alpha_{o},\beta_{o}}$, is itself an
eigenstate of the search Hamiltonian with ${\bf
  H}\ket{\alpha_{o},\beta_{o}} = 0$, and also,
$\braket{\ell}{\alpha_{o},\beta_{o}} = 0$ so that
$\ket{\alpha_{o},\beta_{o}}$ does not contribute to the
time-evolution.  We may also remove eigenstates of $\bf{H}$ which are
at the same time eigenstates of $-\bf{A}$ with the same energy.

This can be seen in the following way. We first consider an
unperturbed eigenstate $\ket{\psi^{o}_{a}}$ such that
$-\mathbf{A}\ket{\psi^{o}_{a}} = E_{a}\ket{\psi^{o}_{a}}$. Let us
assume that there is an eigenvector $\ket{\psi_{a}}$ of the search
Hamiltonian with the same eigenenergy $E_{a}$, that is,
$\mathbf{H}\ket{\psi_{a}} = E_{a}\ket{\psi_{a}}$. Considering the
matrix element $\bra{\psi^{o}_{a}}\mathbf{H}\ket{\psi_{a}}$, we find
$\braket{\psi^{o}_{a} }{ \ell}\braket{\alpha_{o},\beta_{o} }{ \psi_{a}
} + \braket{\psi^{o}_{a} }{ \alpha_{o},\beta_{o}}\braket{\ell }{
  \psi_{a} } = 0$. As $\ket{\alpha_{o},\beta_{o}}$ is an eigenvector
of the search Hamiltonian we know that $\braket{\alpha_{o},\beta_{o}
}{ \psi_{a} } = 0$. This leaves us with $\braket{\psi^{o}_{a} }{
  \alpha_{o},\beta_{o}}\braket{\ell }{ \psi_{a} } = 0$. As the
unperturbed eigenstate $\ket{\psi^{o}_{a}}$ is simply a Bloch state we
know $\braket{\psi^{o}_{a}}{\alpha_{o},\beta_{o}} \neq 0$. Thus, we
obtain $\braket{\ell }{ \psi_{a}} = \sqrt{R_{a}} = 0$. It is then
clear that eigenstates of the search Hamiltonian whose eigenenergies
remain in the spectrum of $-\mathbf{A}$ do not play a role in the
time-evolution of the search.

We now derive an eigenvalue condition for those perturbed eigenvalues
which are in the spectrum of $\bf{H}$.  Using the orthogonality of
eigenstates $\braket{\alpha_{o},\beta_{o}}{\psi_{a}}=0$,
Eq.~(\ref{eq:eigenstate_representation}) leads to
\begin{equation}
  \sqrt{3R_{a}}\bra{\alpha_{o},\beta_{o}}\left(E_{a} + {\bf{A}} \right)^{-1}\ket{\alpha_{o},\beta_{o}} = 0\,.
\end{equation}
By expressing $\bra{\alpha_o,\beta_o}$ in terms of the eigenstates of
the {\em unperturbed } walk Hamiltonian $-{\bf{A}}$, we may write this
as a quantization condition
\begin{equation} \label{FE} F\left(E_{a}\right) = 0
\end{equation}
with
\begin{equation} \label{FE}
  F\left(E\right) =  \frac{\sqrt{3}}{N}
  \sum_{\underline{k}}
  \left[
    \frac{1}{E - \epsilon\left(\underline{k}\right)}
    +
    \frac{1}{E + \epsilon\left(\underline{k}\right)}
  \right]\, ,
\end{equation}
were $N$ is the total number of sites in the lattice.  
Eq.\ (\ref{FE}) is written as two summations to
incorporate the fact that the spectrum of $-\bf{A}$ as well as
$\bf{H}$ is symmetric around $E=0$.

Choosing $\ket{\psi_{a}}$ to be normalised
$\braket{\psi_{a}}{\psi_{a}}=1$,
Eq.~(\ref{eq:eigenstate_representation}) also implies
\begin{equation}
  3 R_{a}\bra{\alpha_{o},\beta_{o}}\left(E_{a} + {\bf A}\right)^{-2}
  \ket{\alpha_{o},\beta_{o}} = 1,
\end{equation}
which allows $R_{a}$ to be rewritten as
\begin{equation}
  R_{a} = \frac{1}{\sqrt{3}|{F'\left(E_{a}\right)}|}.
  \label{eq:Ra}
\end{equation}
We may now rewrite the amplitude in Eq.~\eqref{eq:amplitude} in the
form
\begin{equation}
  \bra{\ell}e^{-i{\bf H}T}\ket{start} = \braket{\alpha_{o},\beta_{o}}{start}\sum_{a}\frac{e^{-iE_{a}T}}{E_{a}|{F'\left(E_{a}\right)}|}\, ,
  \label{eq:time_evolution}
\end{equation}
where we have used the adjoint of
Eq.~(\ref{eq:eigenstate_representation}). Note again that eigenstates
of $\bf{H}$, which are also in the spectrum of $-\bf{A}$, do not
contribute to the time-evolution of the search; it may be seen from
the definition of $F\left(E\right)$ in Eq.~(\ref{FE}) that $\left|
  F'\left(E_{a}\right) \right| \rightarrow \infty$, where $E_{a}$ is
in the spectrum of $-\bf{A}$.

As we saw in the previous section, the avoided crossing is formed by
two perturbers approaching symmetrically either side of the Dirac
states with energy $\epsilon_{D} = 0$.  Thus, we concentrate on
evaluating the contribution to the time-evolution from these perturbed
eigenstates of $\bf{H}$ either side of the Dirac point and we label
these states $\ket{\psi_{\pm}}$.  In order to evaluate these
contributions we calculate the perturbed eigenenergy $E_{+}$ and also
derive a leading-order expression for $F'\left(E_{+}\right)$ (since
the spectrum is symmetric, we have $E_{+} = -E_{-} > 0$).

Using the definition of $F\left(E\right)$ in Eq.~(\ref{FE}), we
estimate $F\left(E_{+}\right)$ by separating out the Dirac points,
where
$\epsilon\left(\underline{K}\right)=\epsilon\left(\underline{K}'\right)
= 0$, from the summations and then Taylor expanding the remaining
terms at $E=0$, to find
\begin{equation}\label{eq:exp}
  F\left(E_{+}\right) = \frac{4\sqrt{3}}{NE_{+}} - \sum_{n=1}^{\infty} I_{2n}E_{+}^{2n-1}\, ,
\end{equation}
where the sums $I_{n}$ are given by
\begin{equation}
  I_{n} = \frac{\sqrt{3}}{N}\sum_{\underline{k} \neq \underline{K},\underline{K}'} 
  \left[
    \frac{1}{\left[\epsilon\left(\underline{k}\right)\right]^{n}}
    + 
    \frac{1}{\left[-\epsilon\left(\underline{k}\right)\right]^{n}}
  \right]
  \,. 
\end{equation}
As the unperturbed spectrum is symmetric only those $I_{n}$ with even
$n$ are non-zero; thus from now on we focus only on $I_{2k}$ where $k\geq
1$.

We stated earlier that the spectrum of graphene is well-approximated
by a conic dispersion relation around the Dirac points.  Thus, it is
from around these points that the major contributions to the $I_{2k}$
sums arise. By applying the linear approximation from
Eq.~(\ref{eq:linear_dispersion}) and the momenta quantum numbers from
Eq.~(\ref{eq:quantised_momenta}) we may approximate the $I_{2k}$
summations as
\begin{eqnarray}
  I_{2k}&= &4\sqrt{3} N^{k-1} \left[ 
    \sum_{(p,q)\in L}
    Z_{2}\left(S_{\underline{K}},2\right)
    + \sum_{(p,q)\in L'}
    Z_{2}\left(S_{\underline{K}'},2\right)
  \right] \nonumber \\ 
  &+& O(1) \, .
  \label{eq:I_epstein}
\end{eqnarray}
Here $Z_2(S,x)$ is the Epstein zeta-function \cite{Siegel}
\begin{equation}
  Z_2(S,x)
  = \frac{1}{2} 
  \sum_{ (p,q) \in \mathbb{Z}^2 
    \backslash (0,0)} 
  \left( S_{11}p^2 
    +2 S_{12} pq + S_{22} q^2
  \right)^{-x}\, ,
\end{equation}
for a real positive definite real symmetric $2 \times 2$ matrix
$S$. For our purposes we use
\begin{equation}
  S_{\underline{K}}=  S_{\underline{K}'}=
  4\pi^2 \begin{pmatrix}
    2 & -1 \\
    -1 & 2
  \end{pmatrix} \, ,
\end{equation}
which describes the spectrum close to the Dirac points. The linear
approximation around both Dirac points $K$ and $K'$ is the same and,
thus, the matrices $S_{\underline{K}}$ and $S_{\underline{K}'}$ are
equal.

Our summation in Eq.~(\ref{eq:I_epstein}) is over the rectangular
regions $L$ and $L'$ of the lattice $\mathbb{Z}^2$. Both are centered
on $(0,0)$ and have side lengths proportional to $\sqrt{N}$, however
the center, $(0,0)$, corresponding to the relevant Dirac point, is
omitted from the summation.

Convergence of the Epstein zeta function is well-known for $k\geq 2$
\cite{Siegel}, and leads to the bounds
\begin{equation}\label{eq:Ikest}
  \lim_{N \to \infty } \frac{I_{2k}}{N^{k-1}}=
  4 \sqrt{3} \left(Z_2(S_{\underline{K}},k)+ Z_2(S_{\underline{K}'},k)
  \right) 
\end{equation}
for $k\ge 2$.
A sharp estimate for
\begin{equation}\label{eq:I2est}
  I_2=  O \left(\ln N\right)\, .
\end{equation}
is given in Appendix~\ref{appendix_a}.  In order to calculate an
estimate for $E_{+}$, we truncate the Taylor expansion of
$F\left(E\right)$ in Eq.~(\ref{eq:exp}) at the $I_{2}$ term (the first
term in the summation), and apply the eigenvalue condition
$F\left(E_{+}\right) = 0$.  That is, we solve $\frac{4 \sqrt{3}}{N
  E_+} - I_2 E_+=0$ and obtain the approximation $E_+^2\approx \frac{4
  \sqrt{3}}{N I_2}$.  This solution also leads to the estimate
$F'\left(E_{+}\right) \approx -2 I_{2}$.

We consider next whether our solution for $E_{+}$ lies inside the
radius of convergence of the Taylor expansion. We note that each term
in $Z_{2}\left(S_{\underline{K}},k\right)$ is smaller than the
corresponding term in $Z_{2}\left(S_{\underline{K}},2\right)$ for
$k>2$, and so it follows that $Z_2(S_{\underline{K}},k)<
Z_2(S_{\underline{K}},2)$ for $k>2$.  This property of the Epstein
zeta function and the estimate from Eq.~(\ref{eq:Ikest}) imply
\[\sum_{n=2}^\infty I_{2n} E_+^{2n-1}< \frac{C}{N
  E_+}\sum_{n=2}^\infty (N E_+^2)^n\] i.e.\ the infinite sum in
Eq.~(\ref{eq:exp}) converges for $E_+< 1/\sqrt{N}$. Thus, for large
$N$, our solution lies within the radius of convergence of our Taylor
expansion in
Eq.~(\ref{eq:exp}).

In order to show that the dominant contributions to the search come
from $\ket{\psi_{\pm}}$, we need to establish the leading order error
term.  All the $I_{2k}$ sums are positive so that, given the sign of
all the terms in the Taylor expansion in Eq.~(\ref{eq:exp}), the true
value of $E_{+} > 0$ has to be smaller than the estimate we have
obtained. Thus, we write the true value of $E_{+}$ as
\begin{equation}
  E_+^2=\frac{4 \sqrt{3}}{N I_2}- \Delta >0\, ,
\end{equation}
with $\Delta> 0$. One may rewrite $F(E_+)=0$, using the true value of
$E_{+}$, to give
\begin{equation}
  I_2 \Delta = \sum_{n=2}^\infty I_{2n} E_+^{2n}\, .
\end{equation}
We follow the same arguments as used for the calculation of the radius
of convergence to obtain an upper bound for the summation. This
together with the already established fact that $E_+$ is inside this
radius for sufficiently large $N$, we get the following inequality:
\begin{eqnarray*}
  0 < N I_2 \Delta &<& C \sum_{n=2}^\infty 
  (N E_+^2)^n\\
  &=& \frac{C N^2 E_+^4}{1- N E_+^2}= O(I_2^{-2}).
\end{eqnarray*}
So $\Delta = O(I_2^{-3}N^{-1})= O\left((\ln N)^{-3}
  N^{-1}\right)$. Thus, we obtain
\begin{equation}
  E_+^2 = \frac{4 \sqrt{3}}{N I_2}\left( 1 + O({(\ln N)^{-2}})\right)\, .
\end{equation}
It also follows that
\begin{equation}
  F'\left(E_{\pm}\right) = -2I_{2} + O\left(\frac{1}{\ln N}\right).
\end{equation}

This shows that our perturbed eigenstates $\ket{\psi_{\pm}}$ have an
$O\left(1\right)$-overlap with the starting state and are, therefore,
the relevant states to be considered in the time-evolution of the
algorithm.  Using the definitions of $\ket{\psi_{a}}$ and $R_{a}$ in
Eqs.~(\ref{eq:eigenstate_representation}), (\ref{eq:Ra}), the inner
product of the starting state and the perturbed eigenvectors can be
expressed as
\begin{equation}
  \braket{start}{\psi_{a}} = \frac{1}{E_a}\sqrt{ \frac{\sqrt{3}}{|F'\left(E_{a}\right)|}}\braket{start}{\alpha_{o},\beta_{o}}\, ,
\end{equation}
where $\braket{start}{\alpha_{o},\beta_{o}}$ is the overlap of the
starting state with the marked vertex state.
Applying our previous approximations for $E_{+}$ and
$F'\left(E_{+}\right)$, that is, for our perturbed eigenstates closest
to the Dirac point, we find
\begin{equation}
  |\braket{start}{\psi_{\pm}}| = \frac{1}{\sqrt{2}} + O\left(\frac{1}{\ln^{2} N}\right).
\end{equation}
Thus, our starting state is indeed a superposition of the perturbed
eigenstates, $\ket{\psi_{\pm}}$ as assumed in the previous section.
This makes it possible to investigate the running time and success
amplitude of the algorithm in more detail, that is,
\begin{align}
  &\left|\bra{\ell}e^{-iHt}\ket{start}\right|  \\
  &\approx \left|\frac{1}{\sqrt{2}}\left(e^{-i E_{+}t}\braket{\ell}{\psi_{+}} - e^{i E_{+}t}\braket{\ell}{\psi_{-}}\right)\right| \\
  &=
  \frac{1}{3^{\frac{1}{4}}I_{2}^{\frac{1}{2}}}\left|\sin\left(E_{+}t\right)\right|\,
  .
\end{align}
It is clear from our earlier results for $E_{\pm}$ and $I_{2}$ that
our algorithm localises on the neighbor state $\ket{\ell}$ in time $T
= \frac{\pi}{2 E_{+}} = O\left(\sqrt{N\ln N}\right)$ with probability
amplitude $O\left(1/\sqrt{\ln N}\right)$.  This confirms the
logarithmic correction observed numerically and displayed in the inset
of Fig.\ \ref{fig:3bond_spectrum}.

\section{Communication}
\label{sec:communication}
It has been demonstrated in \cite{HeinTannerCom} that discrete-time
search algorithms can be modified to create a communication protocol.
We show here that a communication setup can be established also in the
continuous-time search algorithm with minor changes due to the
subtleties of our search.

We use the same unperturbed walk Hamiltonian as before, that is,
${\bf{H_{0}}} = -\bf{A}$ and the same type of perturbation matrix as
in Eq.~(\ref{eq:3bond_perturbation_matrix}), but now at two different
sites in the
lattice. 
Explicitly, our communication Hamiltonian is
\begin{equation} {\bf{H}} = -{\bf{A}} + {\bf{W_{s}}} + {\bf{W_{t}}}\,
  ,
  \label{eq:communication_hamiltonian}
\end{equation}
where the perturbation matrices $\bf{W_{s/t}}$ mark the source and
target sites, respectively.

As prescribed in \cite{HeinTannerCom}, the communication protocol
operates by preparing a state localised on the source perturbation, in
our case on the neighbors of the source site, and allowing the system
to evolve under our communication Hamiltonian.  The system then
localises on the neighboring vertices of the target site.  The effect
of the two perturbation matrices $\bf{W_{s/t}}$ amounts to
disconnecting the two vertices from the underlying lattice.

As noted before, our search algorithm has several different optimal
starting states depending on the position of the marked
vertices. Therefore, we divide our communication analysis into three
different cases: the two sites i) are on the same sublattice and have
the same optimal search starting state; ii) are on the same sublattice
but \emph{do not} share the same optimal search starting state; iii)
are on different sublattices.  Both of the first two cases can be
treated by applying the reduced Hamiltonian method demonstrated
earlier, but communication between different sublattices is not
tractable by this method and so we focus on numerics in this case.

For signal transfer between two sites on the same sublattice, we will
assume, without loss of generality, that our communication takes place
between two sites on the $A$-sublattice,
$\left(\alpha_{s/t},\beta_{s/t}\right)^{A}$; here, the subscript $s$
or $t$ denotes the source or target vertices respectively. Using
arguments as employed in Sec.~\ref{sec:threebondperturbation}, we can
reduce the number of relevant states. Ultimately, the search dynamics
takes place in the subspace spanned by the basis
$\ket{\underline{K}}^{A},\allowbreak\ket{\underline{K}'}^{A}$, the
$A$-type Dirac states, and $\ket{\ell_{s}},\allowbreak\ket{\ell_{t}}$,
the uniform superpositions over the neighbors of the source and
target vertices. This basis leads to the reduced Hamiltonian

\begin{equation}\label{eq:reduced_communication_hamiltonian} {\bf
    \tilde{H}} = \sqrt{\frac{6}{N}}
  \begin{pmatrix}
    0 & 0 & e^{i\mu_{s}} & e^{i\mu_{t}} \\
    0 & 0 & e^{i\nu_{s}} & e^{i\nu_{t}} \\
    e^{-i\mu_{s}} & e^{-i\nu_{s}} & 0 & 0 \\
    e^{-i\mu_{t}} & e^{-i\nu_{t}} & 0 & 0
  \end{pmatrix}\, ,
\end{equation}
where $\mu_{s/t} \equiv \frac{2\pi}{3}\left(\alpha_{s/t} +
  2\beta_{s/t}\right)$ and $\nu_{s/t} \equiv
\frac{2\pi}{3}\left(2\alpha_{s/t} + \beta_{s/t}\right)$.

In the first case considered we assume that there are two
perturbations on the graphene lattice located at the points
$\left(\alpha_{s},\beta_{s}\right)^{A}$ and
$\left(\alpha_{t},\beta_{t}\right)^{A}$, chosen such that
$e^{i\frac{2\pi}{3}\left(\alpha_{s} + 2\beta_{s}\right)} =
e^{i\frac{2\pi}{3}\left(\alpha_{t} + 2\beta_{t}\right)}$. This
implies that a search for either vertex, using our search algorithm
from
Secs.~\ref{sec:threebondperturbation}~\&~\ref{sec:threebondperturbation_analysis},
would use the same optimal starting state. As the phases are equal, we
drop the subscript on the perturbation coordinates in what follows.
Fixing our phases and diagonalising the reduced Hamiltonian, we find
it has eigenvalues $\lambda^{\pm}_{2} = \pm 2\sqrt{\frac{6}{N}}$ and
$\lambda^{1,2}_{0} = 0$ with eigenvectors

\begin{align} 
  \ket{\tilde{\psi}_{\pm 2}} &= \frac{1}{2}\left(e^{i\mu}\ket{\underline{K}}^{A} + e^{i\nu}\ket{\underline{K}'}^{A} \pm \ket{\ell_{s}} \pm \ket{\ell_{t}}\right) \\
  \ket{\tilde{\psi}_{0}^{1}} &= \frac{1}{\sqrt{2}}\left(e^{i\mu}\ket{\underline{K}}^{A} - e^{i\nu}\ket{\underline{K}'}^{A} \right) \\
  \ket{\tilde{\psi}_{0}^{2}} &= \frac{1}{\sqrt{2}}\left(\ket{\ell_{s}}
    - \ket{\ell_{t}}\right)\, .
\end{align}
Using these eigenstates, we may rewrite the source neighbor state as
\begin{equation}
  \ket{\ell_{s}} = \frac{1}{2}\left(\ket{\tilde{\psi}_{2}} - \ket{\tilde{\psi}_{-2}}\right) + \frac{1}{\sqrt{2}}\ket{\tilde{\psi}_{0}^{2}}\, .
\end{equation}
Placing the system in the source state $\ket{\ell_{s}}$ and allowing
the system to evolve under the reduced Hamiltonian, one finds
\begin{align}
  \ket{\psi\left(t\right)} &= e^{-i{\bf \tilde{H}}t}\ket{\ell_{s}} \\
  &= \frac{1}{2}\left(e^{-i\lambda^{+}_{2}t}\ket{\tilde{\psi}_{2}} - e^{-i\lambda^{-}_{2}t}\ket{\tilde{\psi}_{-2}}\right) + \frac{1}{\sqrt{2}}\ket{\tilde{\psi}_{0}^{2}} \\
  &= \frac{-i}{2}e^{i\mu}\sin\left(\lambda^{+}_{2} t\right)\left(\ket{\underline{K}}^{A} + e^{i\frac{2\pi}{3}\left(\alpha -\beta\right)}\ket{\underline{K}'}^{A}\right)  \nonumber \\
  & \hspace{5mm} + \frac{1}{2}\left(\cos\left(\lambda^{+}_{2} t\right)
    + 1\right)\ket{\ell_{s}} +
  \frac{1}{2}\left(\cos\left(\lambda^{+}_{2} t\right) -
    1\right)\ket{\ell_{t}}\, .
\end{align}
We note that, in the last line above, the term in the brackets
involving only the Dirac states is actually the optimal search
starting state for both perturbed vertices, defined in
Eq.~(\ref{eq:starting_state}). The system thus oscillates between the
states localised on the neighbors of the perturbed vertices,
$\ket{\ell_{s}}$ and $\ket{\ell_{t}}$, in a time $T =
\frac{\pi}{2}\sqrt{\frac{N}{6}}$, via their optimal search starting
state.

\begin{figure}[t]
  \centering
  \includegraphics[width =
  1.0\linewidth]{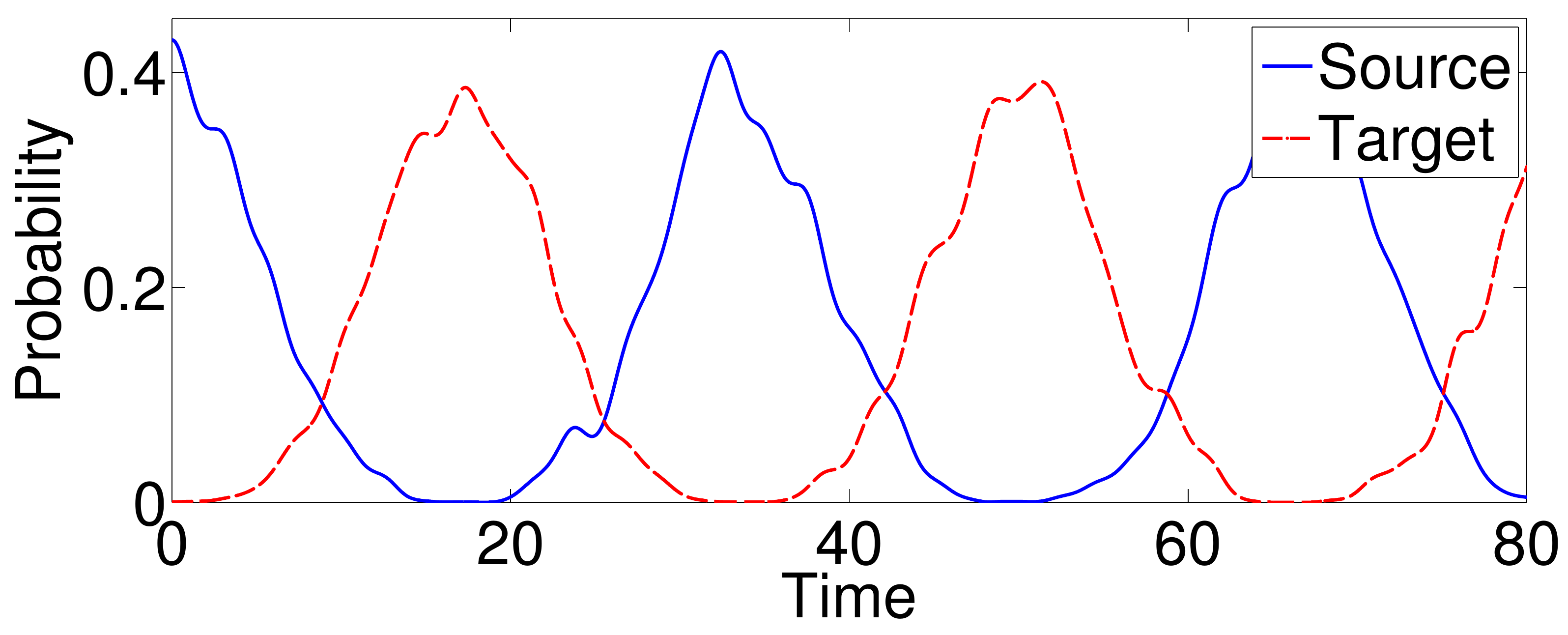}
  \caption{(Color online) Numerically calculated signal transfer on a $12\times 12$
    cell graphene lattice between equivalent vertices, using the
    communication Hamiltonian in
    Eq.~(\ref{eq:communication_hamiltonian}). The system is
    initialised in $\ket{\ell_{s}}$ and localises on
    $\ket{\ell_{t}}$. Only the sum of probabilities to be found on the
    neighbor vertices is shown.}
  \label{fig:equivalent_communication_torus}
\end{figure}

Figure~\ref{fig:equivalent_communication_torus} shows the system
evolving under the full communication Hamiltonian,
Eq.~(\ref{eq:communication_hamiltonian}). The initial state used for
the time evolution shown in
Figure~\ref{fig:equivalent_communication_torus} is the true localised
state, that is, we run the quantum search with a single perturbation
located at vertex $\left(\alpha_{s},\beta_{s}\right)^{A}$ until it
reaches maximum success probability, and then apply the second
perturbation to the vertex
$\left(\alpha_{t},\beta_{t}\right)^{A}$. The figure confirms the
behaviour expected from the reduced model calculation.

The communication mechanism essentially works in the same way as the
quantum search algorithm, as it can be viewed as one marked vertex
`finding' another. The initial localised source state decays back
towards the search starting state, and the system then searches for
the target state.

In our second case of signal transfer we analyse the behaviour of a
communication system where we have two perturbed vertices on the same
sublattice, but with the restriction that
$e^{i\frac{2\pi}{3}\left(\alpha_{s} + 2\beta_{s}\right)} \neq
e^{i\frac{2\pi}{3}\left(\alpha_{t} + 2\beta_{t}\right)}$.  As such,
the two marked sites cannot interact via the same search starting
state. However, the optimal search starting states are not orthogonal
so that signal transfer is still possible, but the resulting
interference effects make the analysis slightly more complicated.

We rewrite the coordinates of the target in terms of the source,
$\alpha_{t} = \alpha_{s} + x$ and $\beta_{t} = \beta_{s} + y$. Again,
fixing our phases and diagonalising the reduced Hamiltonian in
Eq.~(\ref{eq:reduced_communication_hamiltonian}), we find it has
eigenvalues $\lambda^{\pm}_{\sqrt{3}} = \pm\sqrt{3}\sqrt{\frac{6}{N}}$
and $\lambda^{\pm}_{1} = \pm \sqrt{\frac{6}{N}}$ with eigenvectors
\begin{align}
  &\ket{\tilde{\psi}_{\pm\sqrt{3}}} = \frac{e^{i\mu_s}}{2\sqrt{3}}\left(e^{i\frac{2\pi}{3}\left(x + 2y\right)} - 1\right)\ket{\underline{K}}^{A} \\
  &+  \frac{e^{i\nu_s}}{2\sqrt{3}}\left(e^{-i\frac{2\pi}{3}\left(x + 2y\right)} - 1\right)\ket{\underline{K}'}^{A} \nonumber \mp \frac{1}{2}\ket{\ell_{s}} \pm \frac{1}{2}\ket{\ell_{t}} \nonumber \\
  &\ket{\tilde{\psi}_{\pm 1}} = \mp\frac{e^{i\mu_s}}{2}\left(e^{i\frac{2\pi}{3}\left(x + 2y\right)} + 1\right)\ket{\underline{K}}^{A} \\
  &\pm \frac{e^{i\nu_s}e^{i\frac{2\pi}{3}\left(x +
        2y\right)}}{2}\ket{\underline{K}'}^{A} -
  \frac{1}{2}\ket{\ell_{s}} - \frac{1}{2}\ket{\ell_{t}}\, . \nonumber
\end{align}
Using these eigenstates, one may write the source perturbation as
\begin{equation}
  \ket{\ell_{s}} = \frac{1}{2}\left(-\ket{\tilde{\psi}_{\sqrt{3}}} + \ket{\tilde{\psi}_{-\sqrt{3}}} - \ket{\tilde{\psi}_{1}} - \ket{\tilde{\psi}_{-1}}\right)\, . 
\end{equation}
The full expression for the time-evolution is rather cumbersome;
therefore, we only show the terms and prefactors we are interested in,
namely $\ket{\ell_s}$ and $\ket{\ell_t}$
\begin{align}
  \ket{\psi\left(t\right)} = &\frac{1}{2}\left[\cos\left(\lambda^{+}_{\sqrt{3}}t\right)+\cos\left(\lambda^{+}_{1}t\right)\right]\ket{\ell_{s}} \label{eq:simulated_nonequivalent_communication} \\
  &-
  \frac{1}{2}\left[\cos\left(\lambda^{+}_{\sqrt{3}}t\right)-\cos\left(\lambda^{+}_{1}t\right)\right]\ket{\ell_{t}}
  \nonumber \, .
\end{align}
We can see here that the prefactors do not depend upon the coordinates
of either the source or the target sites; the transport signal between
sites on the same sublattice but with different optimal search
starting state is thus independent of the position of the source or
target site.

\begin{figure}[t]
  \centering
  \includegraphics[width =
  1.0\linewidth]{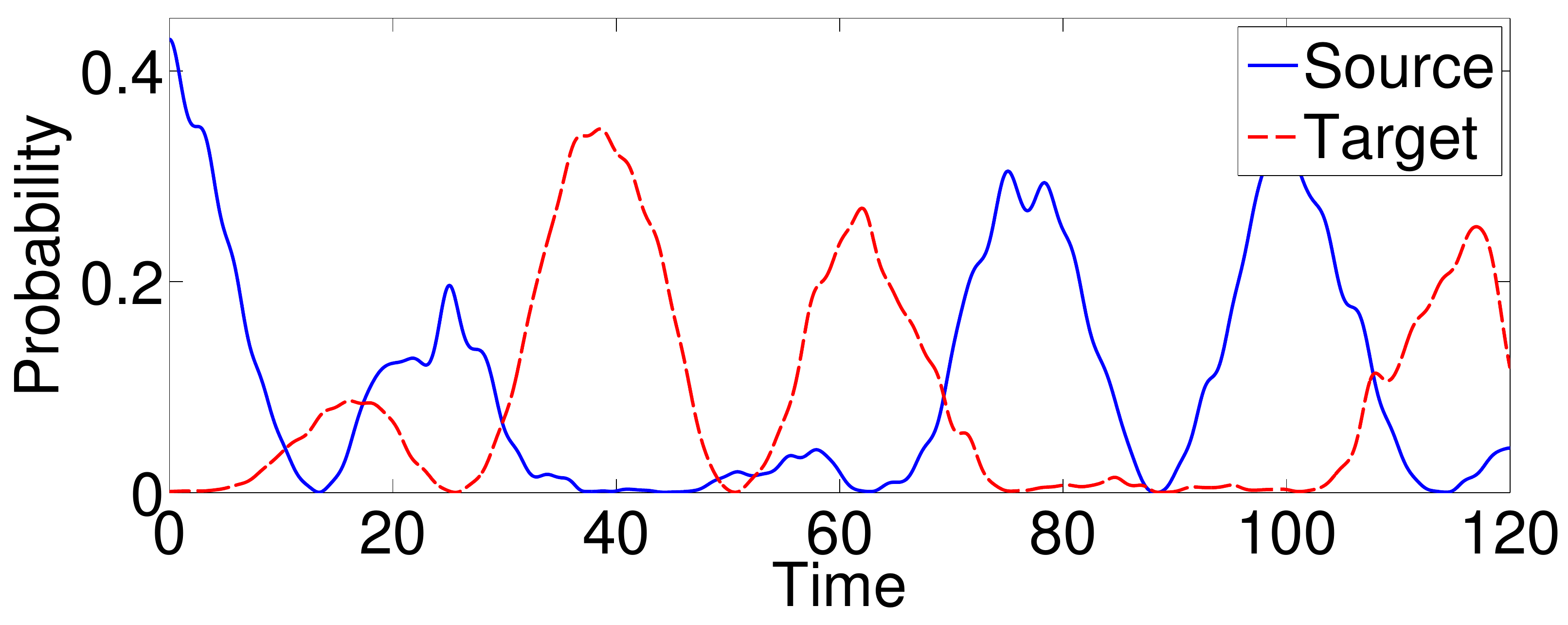}
  \newline \centering
  \includegraphics[width =
  1.0\linewidth]{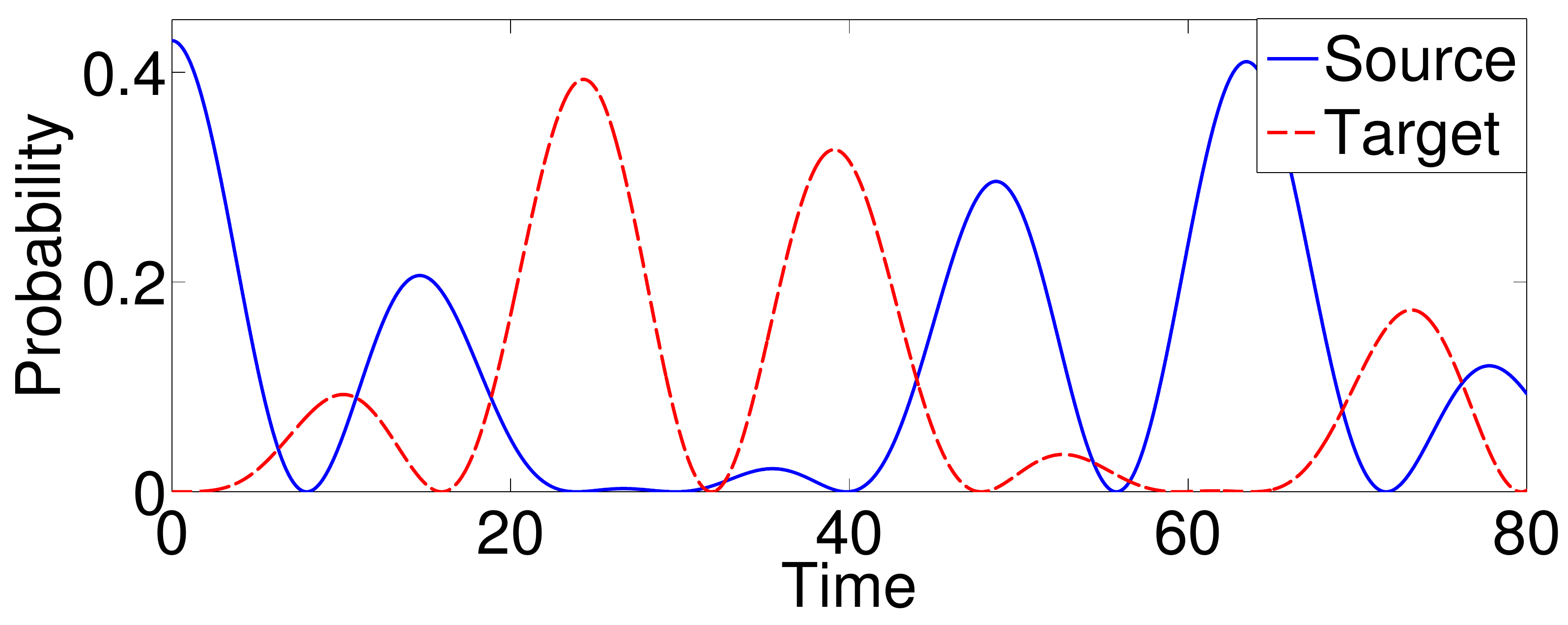}
  \caption{(Color online) Upper: Numerically calculated signal transfer on a
    $12\times 12$ cell graphene lattice between non-equivalent
    vertices, using the communication Hamiltonian in
    Eq.~(\ref{eq:communication_hamiltonian}). The system is
    initialised in $\ket{\ell_{s}}$ and localises on
    $\ket{\ell_{t}}$. Only the sum of probabilities to be found on the
    neighbor vertices is shown. Lower: Analytically calculated
    behaviour for the same system, using the reduced Hamiltonian
    method and Eq.~(\ref{eq:simulated_nonequivalent_communication}).}
  \label{fig:both_nonequivalent_communication}
\end{figure}

In Figure~\ref{fig:both_nonequivalent_communication} we show the
system evolved under the full communication Hamiltonian.  Again, the
initial state is the true localised state on the nearest-neighbors of
the source vertex, obtained by running the search algorithm using one
marked vertex until it reaches its peak success probability. The
time-evolution is radically different to the previous communication
case; the behaviour here is erratic with uneven peaks of probability
at the two perturbations involved in the protocol. However, there are
still significant probability revivals.

The transport behaviour from
Eq.~(\ref{eq:simulated_nonequivalent_communication}), calculated using
the reduced Hamiltonian, is also shown in
Figure~\ref{fig:both_nonequivalent_communication}. The probability at
time $t = 0$ has been scaled to match that shown in
Figure~\ref{fig:both_nonequivalent_communication}. Our calculated
behaviour has the same signal pattern as the numerically calculated
behaviour from the full Hamiltonian, although over a shorter period of
time. As our reduced model only makes use of the Dirac states and the
perturber states, we lose the contribution to the time-evolution from
the rest of the spectrum giving rise to logarithmic corrections as
discussed in Sec.~\ref{sec:threebondperturbation_analysis}. This leads
to differences in the overall time scales for the reduced model and
the full Hamiltonian. This also supports our findings that signal transfer
between all non-equivalent vertices on the same sublattice is the
same. The communication protocol is again set up by a 
search mechanism in reverse, where one vertex finds another. The
slightly erratic behaviour that emerges here is due to interference
between the two separate search mechanisms which interact due to the
non-zero inner product of the three possible optimal search starting
states for vertices on the $A$-sublattice.

\begin{figure}[t]
  \centering
  \includegraphics[width =
  1.0\linewidth]{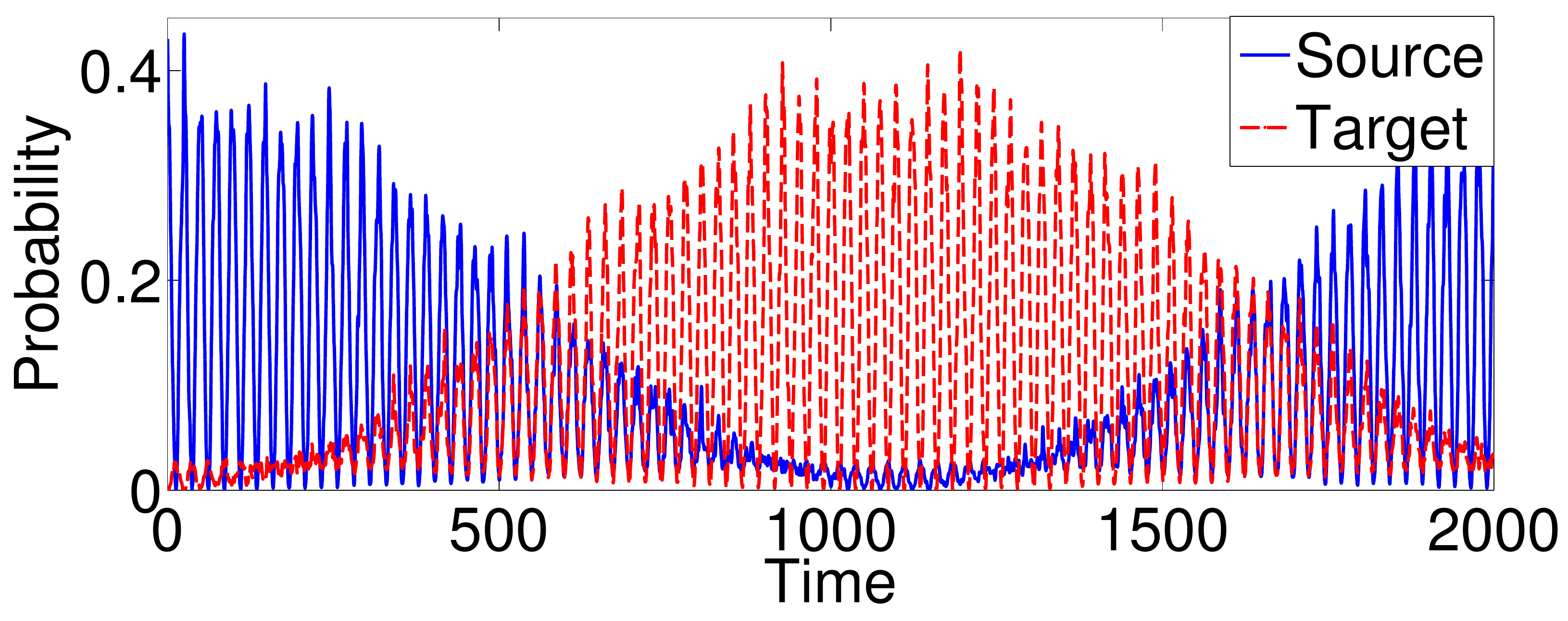}
  \newline \centering
  \includegraphics[width =
  1.0\linewidth]{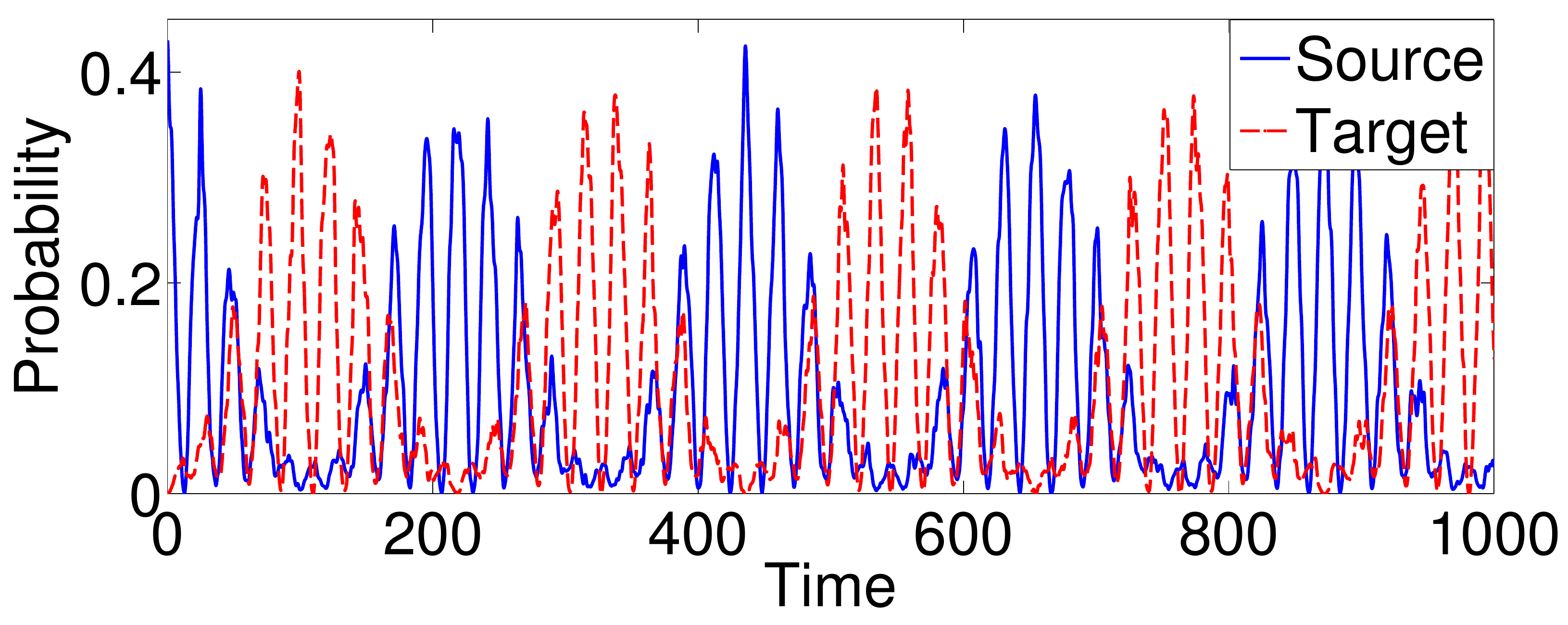}
  \caption{(Color online) Numerically calculated signal transfer on a $12\times 12$
    cell graphene lattice between vertices on different sublattices,
    using the communication Hamiltonian in
    Equation~\ref{eq:communication_hamiltonian}. The system is
    initialised in $\ket{\ell_{s}}$ and localises on
    $\ket{\ell_{t}}$. Only the sum of probabilities to be found on the
    neighbor vertices is shown. The two figures have the same source
    position but different positions of the target site.}
  \label{fig:both_differentsublattice_communication}
\end{figure}

In our final case of signal transfer, we consider communication
between sites on different sublattices which we can not treat in the reduced
Hamiltonian model. It has been demonstrated previously that
perturbations to one sublattice do not interact with the Dirac states
which live on the other sublattice.  Therefore, when attempting to
reduce the communication Hamiltonian as done before, we merely find
that it decouples into two non-interacting reduced Hamiltonians, each
describing a search protocol on one sublattice. The interaction
between the two sublattices is here facilitated due to interactions
via the bulk of the spectrum. In what follows, we focus on numerics
and inspect the behaviour for systems involving the same source
perturber but different targets.  The numerics show a finite number of
different signal patterns, two examples are shown in
Figure~\ref{fig:both_differentsublattice_communication}. We can see
from these examples that the communication mechanism takes place over
a much longer timescale with a superimposed oscillatory dynamics.

The increase in timescale can be attributed to the weak nature of the
interaction between the two perturbations due the fact that the
localised states live mainly on one sublattice. The signal pattern in
Figure~\ref{fig:both_differentsublattice_communication} thus decouples
into a fast oscillation between one site, say the source site, and the
delocalised lattice state and a slow time scale on which a small
amount of probability amplitude escapes into the other sublattice due
to the weak interaction. This process continues until the recurrence
probability at the target perturbation reaches the same peak as the
initial localised source state, and then the behaviour reverses.

\section{Alternative methods of marking}
\label{sec:alternativemarking}
In the previous sections, we have discussed a method of marking a
vertex through modifying the hopping potential from the perturbed site
to all three of its nearest-neighbors.  Here, we will discuss
alternative methods of marking a vertex which still keep the perturber
interaction in the conic dispersion region of the spectrum. This can
be done by altering the hopping potential in different ways as
discussed above; this approach is necessarily a rank-2 perturbation
and these two perturber states must meet at the Dirac energy. We will
discuss several ways of applying such a perturbation in the next
paragraph.  Another approach is based on coupling extra sites to the
lattice; this set-up will be discussed at the end of this section.

Focussing on hopping potential perturbations, we may consider
perturbing the bonds to any number of nearest-neighbors to any
strength. We have seen previously that a symmetric three-bond
perturbation successfully creates a search. Here, we demonstrate a
search based on perturbing the hopping potential from a given site to
only one of its nearest-neighbors. That is, we use the same search
Hamiltonian as in Eq.~(\ref{eq:full_search_Hamiltonian}) with the
perturbation matrix
\begin{equation} {\bf{W}} =
  \ket{\alpha_{o},\beta_{o}}^{A}\bra{\alpha_{o},\beta_{o}}^{B} +
  \ket{\alpha_{o},\beta_{o}}^{B}\bra{\alpha_{o},\beta_{o}}^{A}\, ,
  \label{eq:singlebond_perturbation}
\end{equation}
and eigenstates
\begin{align}
  \ket{W_{g}} &= \frac{1}{\sqrt{2}}\left(\ket{\alpha_{o},\beta_{o}}^{A} -\ket{\alpha_{o},\beta_{o}}^{B}\right) \\
  \ket{W_{e}} &=
  \frac{1}{\sqrt{2}}\left(\ket{\alpha_{o},\beta_{o}}^{A}
    +\ket{\alpha_{o},\beta_{o}}^{B}\right).
\end{align}
Note that it no longer makes sense to speak of a single marked
vertex. Rather, our perturbation marks both vertices
$\left(\alpha_{o},\beta_{o}\right)^{A/B}$ simultaneously; thus, it may
be more accurate to view our perturbation matrix as marking a single
cell of the lattice.

\begin{figure}
  \begin{overpic}[width = 1.05\linewidth]{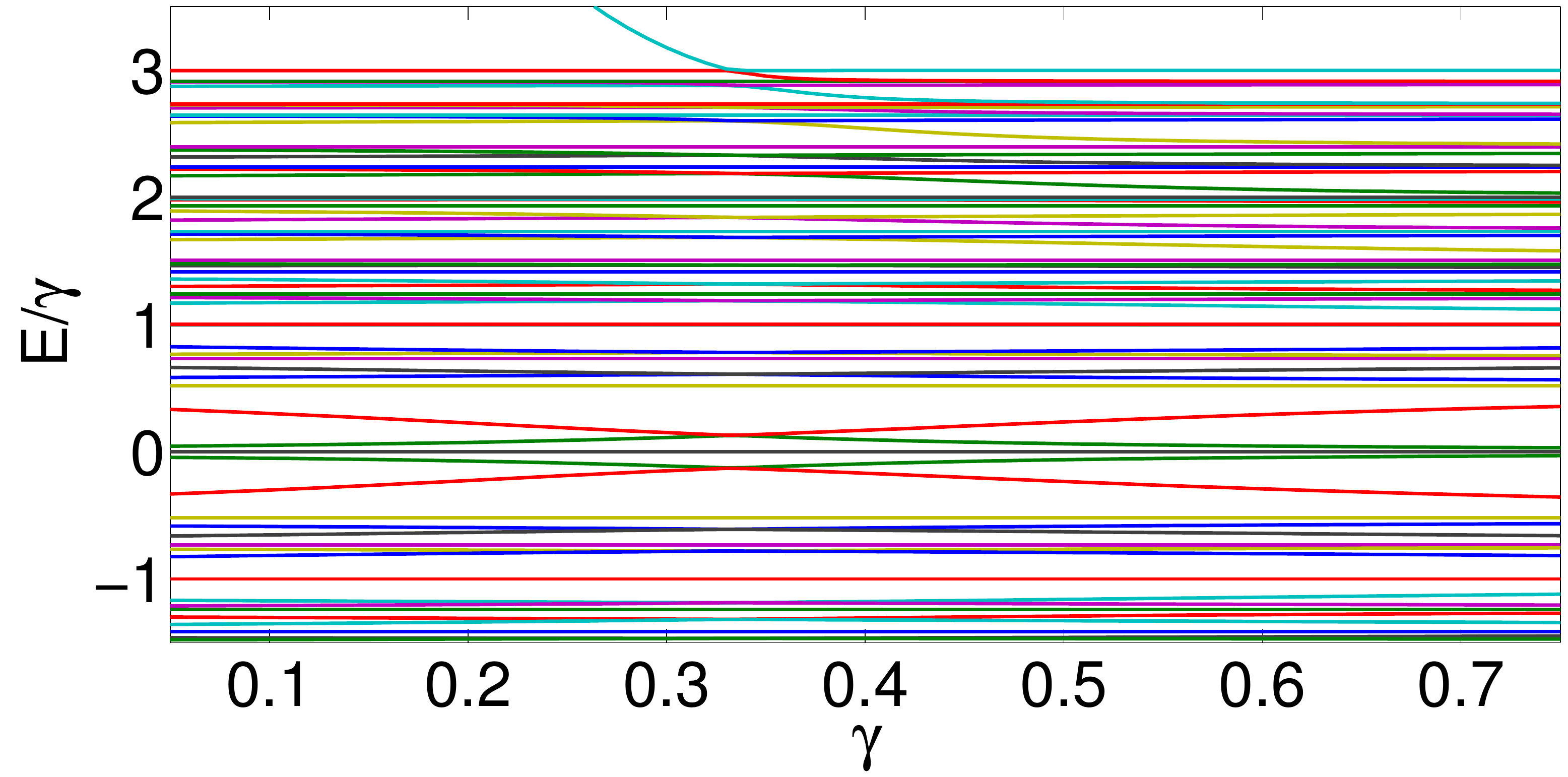}
    \put(54.26,26.75){\fbox{\colorbox{white}{\includegraphics[width=0.45\linewidth]{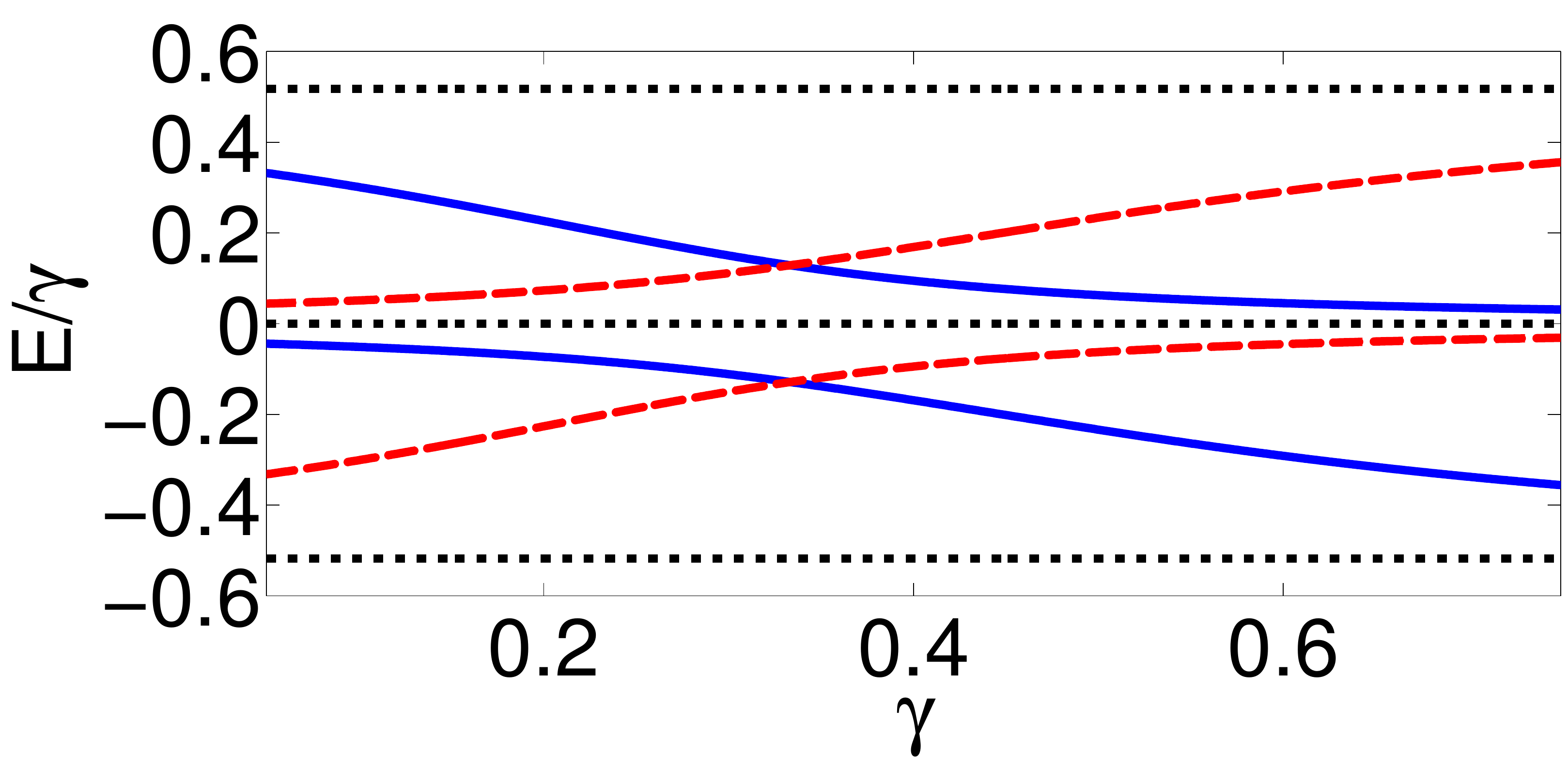}}}}
  \end{overpic}
  \caption{(Color online) Spectrum of single-bond search Hamiltonian in
    Eq.~(\ref{eq:full_search_Hamiltonian}) using perturbation from
    Eq.~\ref{eq:singlebond_perturbation} as a function of $\gamma$ for
    a $12\times 12$ cell torus ($N=288$). The spectrum is symmetric
    around $\epsilon_{D} = 0$.  Inset: States nearest the Dirac energy
    have been coloured depending on their parity with respect to the
    $C_{2}$ operator: even (solid blue), odd (dashed red), undefined (dot-dashed black).}
  \label{fig:singlebond_spectrum}
\end{figure}

For this type of perturbation, the avoided crossing used to generate
search behaviour is not necessarily at $\gamma = 1$, see
Figure~\ref{fig:singlebond_spectrum}, where the spectrum of our search
Hamiltonian as a function of $\gamma$ for a $12\times12$ cell torus is
shown. As stated before, our single bond perturbation is also a rank-2
matrix, and so again there are two perturber states approaching the
spectrum from the negative and positive regions of the
spectrum. Inspecting the region around the Dirac energy, see inset in
Figure~\ref{fig:singlebond_spectrum}, we see that avoided crossings
are formed by four states in total, the two blue (solid) curves and the two
red (dashed) curves. At $\gamma = \frac{1}{3}$ there is an exact crossing
between the red and blue curves indicating that these states are
orthogonal. 

We can make this avoided crossing picture clearer by breaking down the
eigenstates of the search Hamiltonian in terms of the symmetries of
the lattice. In particular, we use the rotation operator $C_2$, which
is a rotation of $\pi$ about the mid-point of the perturbed
bond. Considering the action of $C_{2}$ on the marked
vertices and the eigenstates of the perturbation matrix, one finds
$C_{2}\ket{\alpha_{o},\beta_{o}}^{A/B} =
\ket{\alpha_{o},\beta_{o}}^{B/A}$, $C_{2}\ket{W_{g}} = -\ket{W_{g}}$
and $C_{2}\ket{W_{e}} = \ket{W_{e}}$.  The inset of
Figure~\ref{fig:singlebond_spectrum} shows the states nearest to the
Dirac energy, where the same colour indicates the same $C_{2}$
parity. It now becomes clear that the states near the Dirac point
actually form two avoided crossings, one for each parity with a
minimum energy gap at $\gamma = \frac{1}{3}$.

\begin{figure}[t]
  \centering
  \includegraphics[width = 1.0\linewidth]{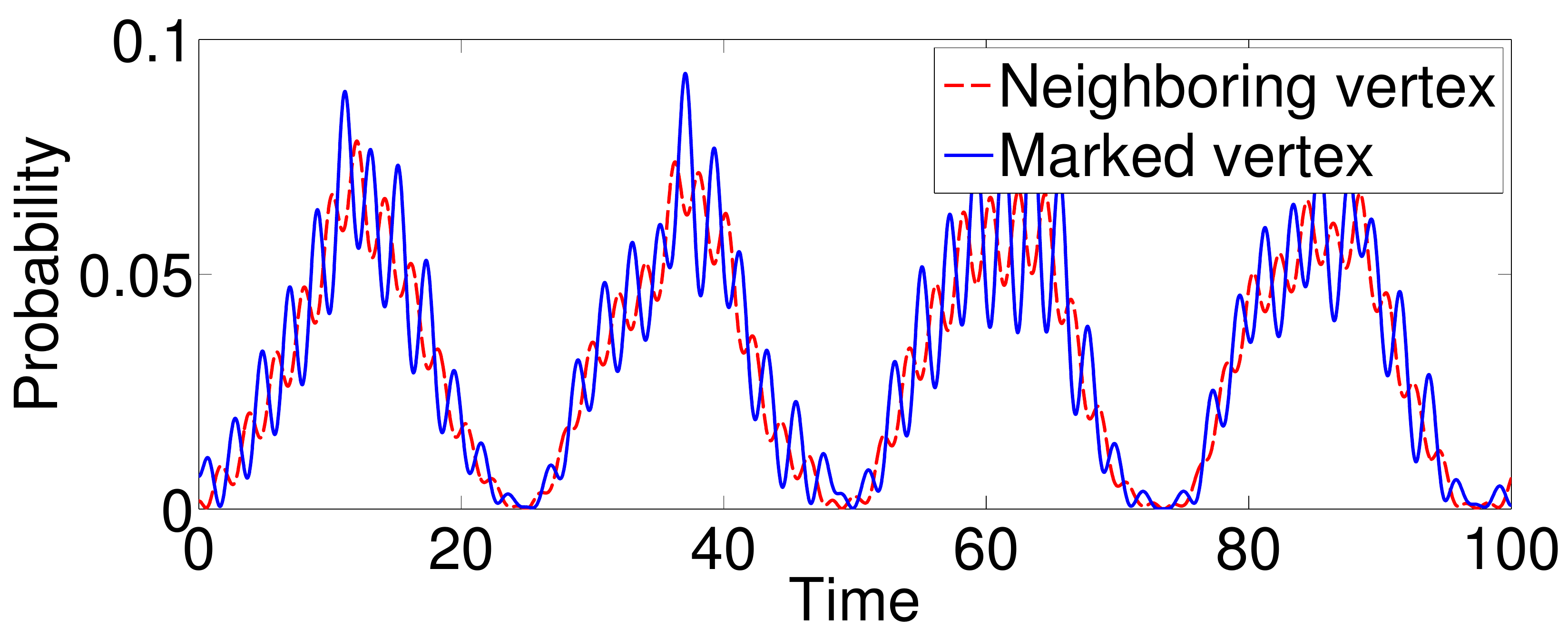}
  \caption{(Color online) Search on $12\times 12$ cell graphene lattice with
    single-bond perturbation, using an optimal starting state. The
    behaviour at both marked vertices is the same, as is the behaviour
    at each of their neighboring sites.}
  \label{fig:singlebond_search}
\end{figure}
Without going into details, we can again find optimal starting states
using a reduced Hamiltonian approach (one for each of the avoided crossings); the evolution of the system
using these optimal states is shown in
Figure~\ref{fig:singlebond_search} for a $12\times12$ cell lattice.
We find there is a significant localisation on the marked vertices and
their nearest-neighbors, with the probability of being found on
either of the marked vertices peaking at around $16-18\%$.  We can
also see that the probability of being found on each of the
nearest-neighbors peaks at around $8\%$, resulting in a total
probability of being found on the marked vertices and their
nearest-neighbors of approximately $48\%$.
The success probability fluctuates, as the probability amplitude
oscillates between the marked vertices and their
nearest-neighbors. This is due to the probability amplitude being
constrained in the local area by the increased hopping potential
between the two marked vertices.

Although our demonstration is for a $12\times12$ cell lattice,
numerical investigations show that the search behaviour remains the
same as the lattice size increases with critical value fixed at
$\gamma = \frac{1}{3}$. We also find that the gap at the two avoided
crossings scales as $\mathcal{O}\left(1/\sqrt{N\ln N}\right)$, in the
same way as for the three-bond perturbation shown in
Figure~\ref{fig:3bond_spectrum}. As the search time is inversely
proportional to the energy splitting at the avoided crossing, it also
gives an estimate of the running time of the search $T =
\mathcal{O}\left(\sqrt{N\ln N}\right)$.

We now turn to coupling additional sites to the lattice as a way of
introducing a perturbation as also considered in \cite{Nice}; such a treatment 
is in many ways closer to
experimental realisations as defects due to additional add-on atoms are
quite common in graphene \cite{Banhart}. We focus here on the
idealised case where an additional site is coupled to a single lattice
vertex. We use the perturbation matrix
\begin{equation} {\bf{W\left(\gamma\right)}} =
  -\ket{\alpha_{o},\beta_{o}}^{A}\bra{site}\, -\,
  \ket{site}\bra{\alpha_{o},\beta_{o}}^{A} +
  \gamma\ket{site}\bra{site}\, ,
\end{equation}
for coupling the additional site $\ket{site}$ to the $A$-type vertex
$\left(\alpha_{o},\beta_{o}\right)^{A}$ and $\gamma$ is a free
parameter related to the on-site energy of the additional vertex. The
choice of the coupling terms is such that the binding energy between
the additional site and the lattice vertex is the same as the internal
couplings in the lattice. Thus, our search Hamiltonian is of the form
\begin{equation} {\bf{H}} = -{\bf{A}} + {\bf{W}\left(\gamma\right)}\,
  .
  \label{eq:additional_site}
\end{equation}

\begin{figure}
  \begin{overpic}[width = 1.05\linewidth]{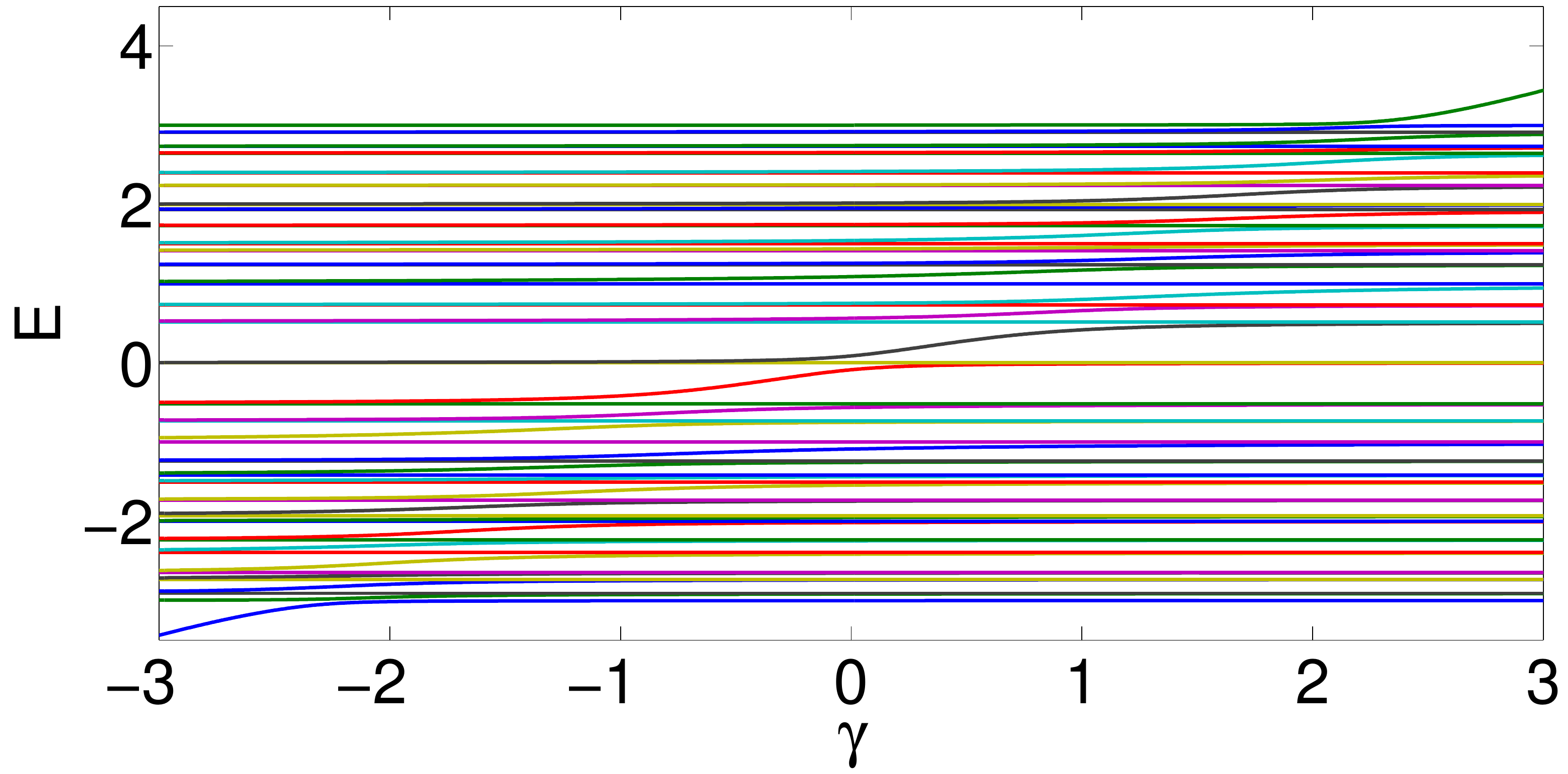}
    \put(11.75,28.25){\fbox{\colorbox{white}{\includegraphics[width=0.45\linewidth]{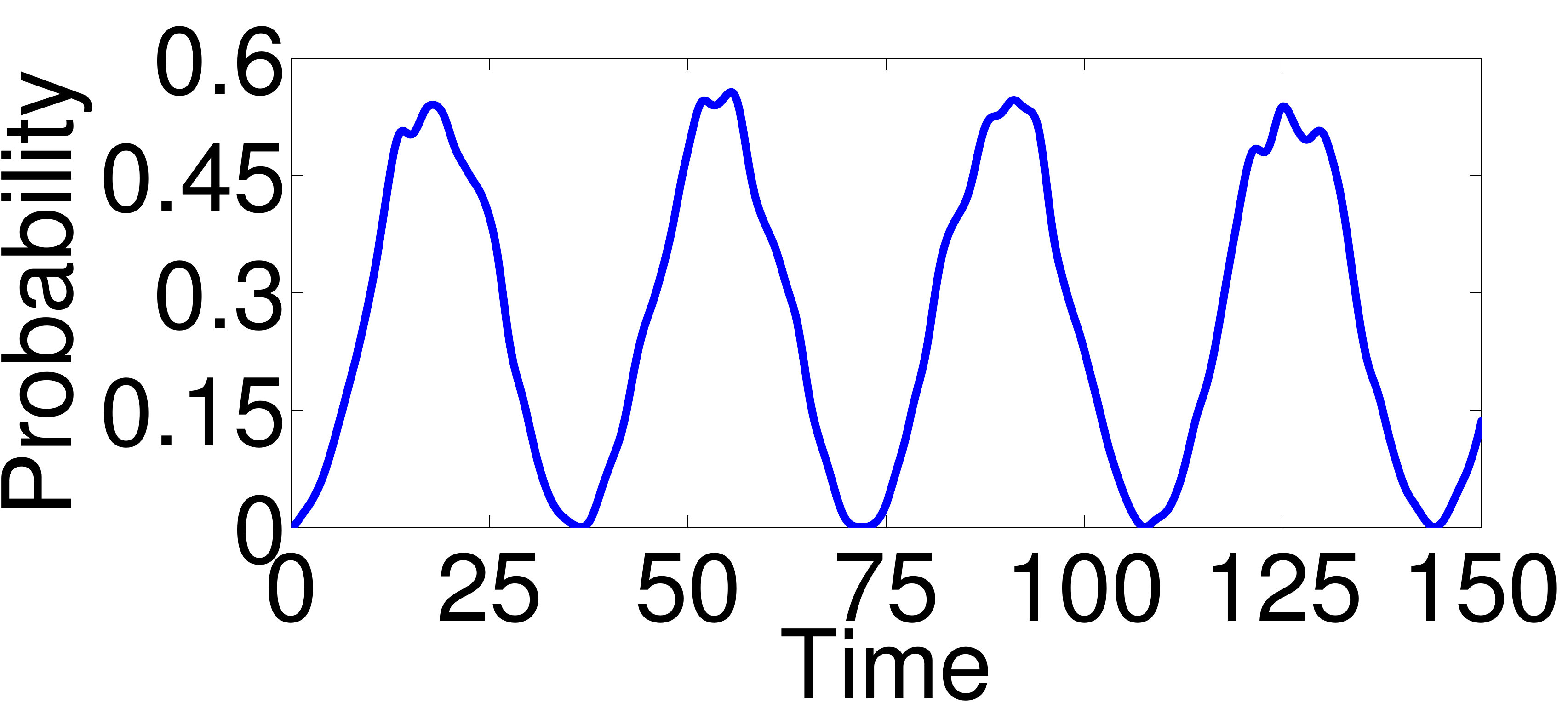}}}}
  \end{overpic}
  \caption{(Color online) Spectrum of the Hamiltonian in
    Eq.~(\ref{eq:additional_site}) as a function of $\gamma$ for a
    $12\times 12$ cell torus ($N=288$). The symmetry of the graphene
    spectrum is broken by the choice of the perturbation in terms of
    an additional site. Inset: Search on $12\times 12$ cell graphene
    lattice with a single additional site perturbation, using starting
    state $\ket{s}$ from Eq.~(\ref{eq:starting_state})}
  \label{fig:singlesite_spectrum}
\end{figure}

In Figure~\ref{fig:singlesite_spectrum}, the spectrum of the search
Hamiltonian is given as function of $\gamma$. As our free parameter
$\gamma$ only changes a single term, our spectrum only has a single
perturber state. One finds a clear avoided crossing around the Dirac
energy ($E=0$) when $\gamma =0$, that is, when the on-site energy of
the additional site matches the on-site energy of the lattice
vertices. 

In the reduced Hamiltonian picture, we write
Eq.~\ref{eq:additional_site} in terms of a basis consisting of the
Dirac states and our perturber state, $\ket{site}$. Only three states
are involved in the search, that is,
$\{\ket{\underline{K}}^{A},\ket{\underline{K}'}^{A},\ket{site}\}$, and
$\bf H$ reduces, up to a different prefactor, to the same $3\times3$
matrix found in Eq.~(\ref{redH}) for the three-bound
perturbation. 
Thus, our previous reduced Hamiltonian analysis holds for this case
and the optimal search starting states are the same; the
time-evolution for the additional site search is shown in the inset in
Figure~\ref{fig:singlesite_spectrum}.  The only differences between
this case and the three-bond marking are that the system localises on
the additional site, and also the change in prefactor leads to a
search time of $T = \frac{\pi}{4}\sqrt{N}$.

We note that other, more realistic types of perturbations involving
additional sites coupling to a lattice site and its neighbors, can be
shown to result in effective search
protocols.  
Also, the single additional site perturbation described here can, like
the three-bond perturbation, be used to setup a communication
protocol, with the same signal patterns and behaviours as found
previously. The initial state is in this case completely localised on
the additional site.

\section{Alternative nanostructures}
\label{sec:alternativenanostructures}

So far we have described the dynamics of searches on graphene
lattices with periodic boundary conditions, that is, graphene on a
torus.  Here we consider more realistic boundary conditions such
as nanotubes and graphene sheets.

\begin{figure}[t]
  \centering
  \includegraphics[width =
  0.8\linewidth]{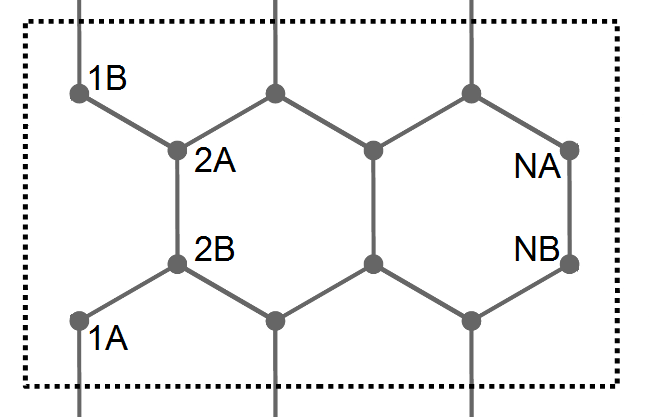}
  \caption{Example of a armchair nanotube cell.  The nanotube is
    periodic along the vertical axis and finite in the horizontal
    direction with a width of $N_{x}$ sites. }
  \label{fig:armchairnanotube_example}
\end{figure}

For the dynamics on nanotubes, we move to periodic boundary condition
along one axis only and impose Dirichlet boundary conditions along the
other directions. The properties of nanotubes are well-known
\cite{Wong}, and it has been shown that the band structure of armchair
nanotubes, that is, nanotubes with armchair boundaries, always allows 
for an energy at the Dirac energy regardless
of the nanotube diameter; we will focus on these types of nanotubes.
We are interested in searching on finite length nanotubes, where the
band structure becomes discrete; the length of the nanotube are in
addition chosen such that there exists an eigenenergy at the Dirac
energy. An example of the cell we use to construct finite armchair
nanotubes is shown in Figure~\ref{fig:armchairnanotube_example}. We
choose the finite length of the nanotube to be along the horizontal
axis and we close the underlying graphene lattice into a nanotube
along the vertical axis.

We denote the basis states of sites in our nanotube as
$\ket{m,A/B,l}$, where $m$ indicates the $m^{th}$ $A/B$-type vertex in
the horizontal direction in the $l^{th}$ cell. It is simple to see
that, due to the periodic boundary conditions along the circumference
of the tube, the vertical component of the eigenstates must be Bloch
states and the horizontal component of the amplitudes must be
sinusoidal in nature. That is, the eigenstates of the finite nanotube
are standing waves along its length.

By working through the tight-binding model for this system, one can
show that $\left(k_{x},k_{y}\right) = \left(\frac{2\pi}{3},0\right)$
is the only point where there exists an eigenenergy equal to the Dirac
energy. It is also possible to show that the spectrum includes Dirac
points when the number of sites along the length of the nanotube
$N_{x} = 3r -1$, where $r$ is an integer.

As there is only one potential Dirac point for finite armchair
nanotubes, it follows that there are only two Dirac states (one from
the bonding and the anti-bonding regions of the spectrum). As we have
$k_{y} = 0$ at the Dirac point, the Bloch wave around the
circumference of the nanotube is simply a uniform
superposition. Another important feature of the Dirac states on the
nanotube, which we have not encountered previously, is the existence
of nodal points where the amplitude of the eigenstate is 0. We find
these nodal points occur at every third site along the horizontal
axis.

We focus on applying our three-bond perturbation from
Sections~\ref{sec:threebondperturbation}~\&~\ref{sec:threebondperturbation_analysis}
to the armchair nanotube. In our new labelling our perturbation takes
the form
\begin{align}
  W = &\ket{m_{o},A,l_{o}}\left(\bra{m_{o}+1,B,l_{o}} +
    \bra{m_{o}-1,B,l_{o}} \right. \nonumber \\ &\left. +
    \bra{m_{o},B,l_{o}}\right) + h.c.\, .
\end{align}
We have assumed that we are perturbing an $A$-type vertex on an even
horizontal coordinate, so that the perturbed site and its
nearest-neighbors remain within one nanotube cell. While it is easy
to re-write this perturbation matrix for other sites we restrict
ourselves to this form for simplicity. Note even in this form it can
still be expressed in terms of a marked site, $\ket{m_{o},A,l_{o}}$,
and a state which lives on the neighbors, $\ket{\ell} =
\frac{1}{\sqrt{3}}\left(\ket{m_{o}+1,B,l_{o}} + \ket{m_{o}-1,B,l_{o}}
  + \ket{m_{o},B,l_{o}}\right)$.

Our search Hamiltonian for the nanotube is the same as in
Eq.~(\ref{eq:full_search_Hamiltonian}). Inspecting the spectrum of the
search Hamiltonian as a function of the free parameter $\gamma$ where
the perturbation is not located on a nodal point of a Dirac state, one
finds an avoided crossing around the Dirac energy when $\gamma = 1$.
This spectrum is very similar to the one shown in
Figure~\ref{fig:3bond_spectrum} for the search on the torus and will
be
omitted. 
If the perturbation is located on a nodal site the picture is
different, however; one finds that there is an exact crossing at the
Dirac energy when $\gamma = 1$ and searching is not
possible.  
Recall that the effect of the perturbation at the critical point is to
completely remove the site from the lattice, if the site already has
zero amplitude then the perturbation will not interact with the Dirac
state.

Similar to the previous section, we numerically reduce the full search
Hamiltonian in a basis consisting of the two Dirac states and the
state living purely on the neighbors, $\ket{\ell}$. The perturbed
site, $\ket{m_{o},A,l_{o}}$, is decoupled from the lattice.  Through
this process, we find two possible starting states, one for each
sublattice.  The starting states are weighted superpositions of the
Dirac states over the sublattice containing the marked vertex. The
nodal points are excluded from this treatment.

\begin{figure}[t]
  \centering
  \includegraphics[width =
  1.0\linewidth]{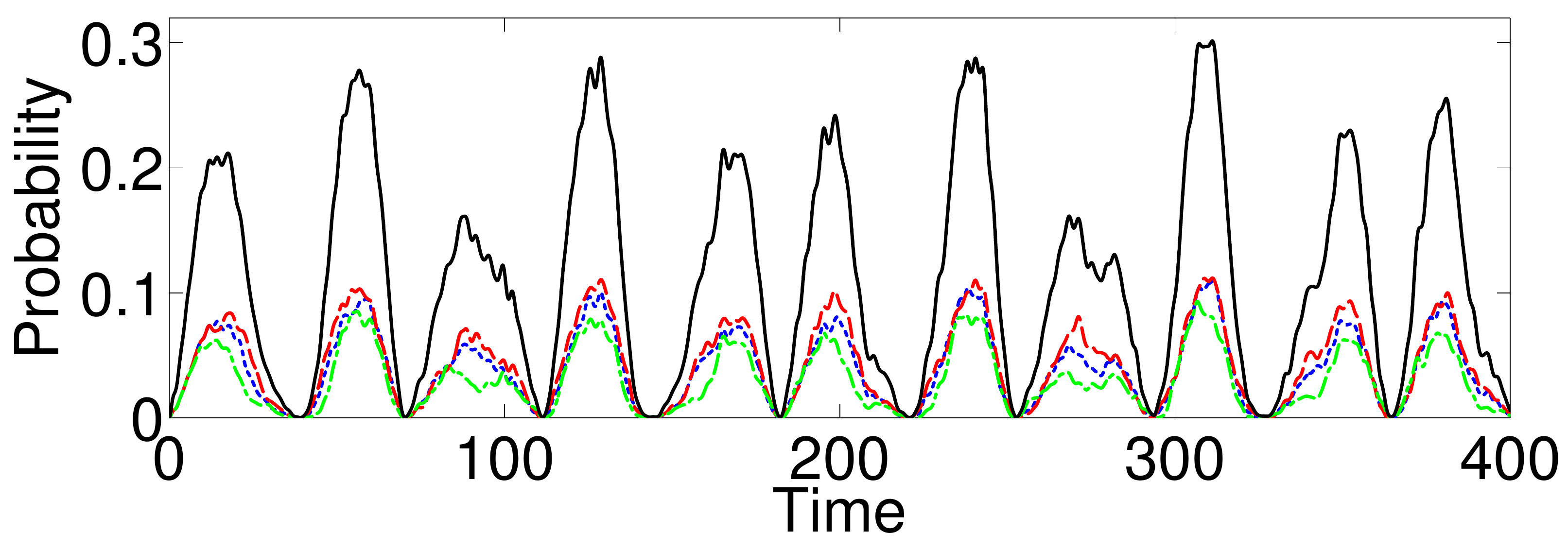}
  \newline \centering
  \includegraphics[width =
  1.0\linewidth]{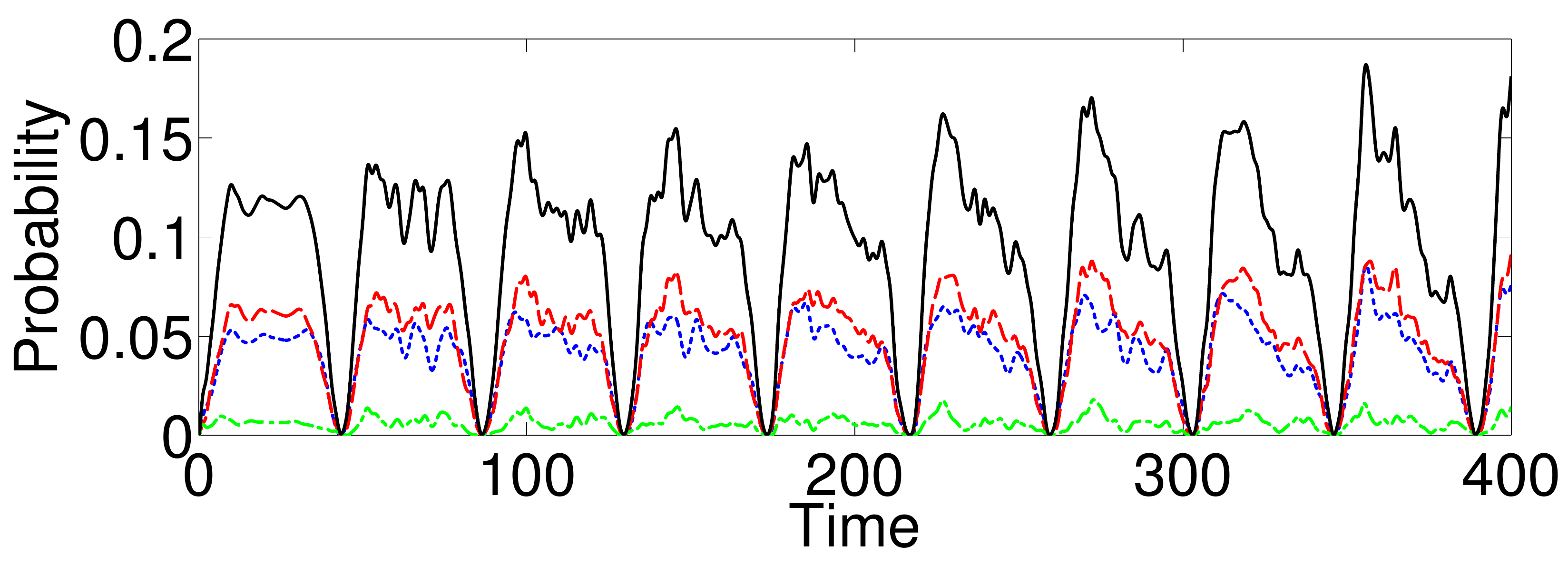}
  \caption{(Color online) Searches on an armchair nanotube ($N=320$) with triple-bond
    perturbation located in the bulk (upper figure) and placed near
    the edge of the nanotube(lower figure), using numerically
    calculated optimal starting state: $\ket{m_{o}-1,B,l_{o}}$ (dotted blue),
    $\ket{m_{o}+1,B,l_{o}}$ (dashed red), $\ket{m_{o},B,l_{o}}$ (dot-dashed green), sum
    of neighbor probabilities (solid black).}
  \label{fig:armchair_search}
\end{figure}

Figure~\ref{fig:armchair_search} shows the system evolution for two
searches using the numerically found optimal starting states. One
search has the marked site located in the centre of the nanotube, the
other has the marked site positioned near the edge of the nanotube.
The dimensions of the nanotube have been chosen so that there are 320
sites, comparable to the searches shown in earlier sections. Comparing
these searches to the three-bond perturbation search on the torus in
Figure~\ref{fig:search_prob}, we see that there is a marked difference
in behaviour and success probability induced by relaxing the periodic
boundary conditions along one axis.

Using a perturbation located near the edge, the lower image in
Figure~\ref{fig:armchair_search}, we see a reduction in success
probability by a factor of 2-3 and strong fluctuations in peak height
when compared to the search on the torus.  Searching in the bulk,
shown in the lower image of Figure~\ref{fig:armchair_search}, displays
behaviour closer to searches on a graphene torus, but again with
significant fluctuations at each peak. We propose that this effect is
due to the reflection of probability amplitude from the edges of the
nanotube. This is supported by the changes in the interference pattern
in the signal as the perturbation is moved across the lattice.
Numerics demonstrate that additional site perturbations and
communication protocols can also be used on nanotube
lattices. 

\begin{figure}[t]
  \centering
  \includegraphics[width =
  1.0\linewidth]{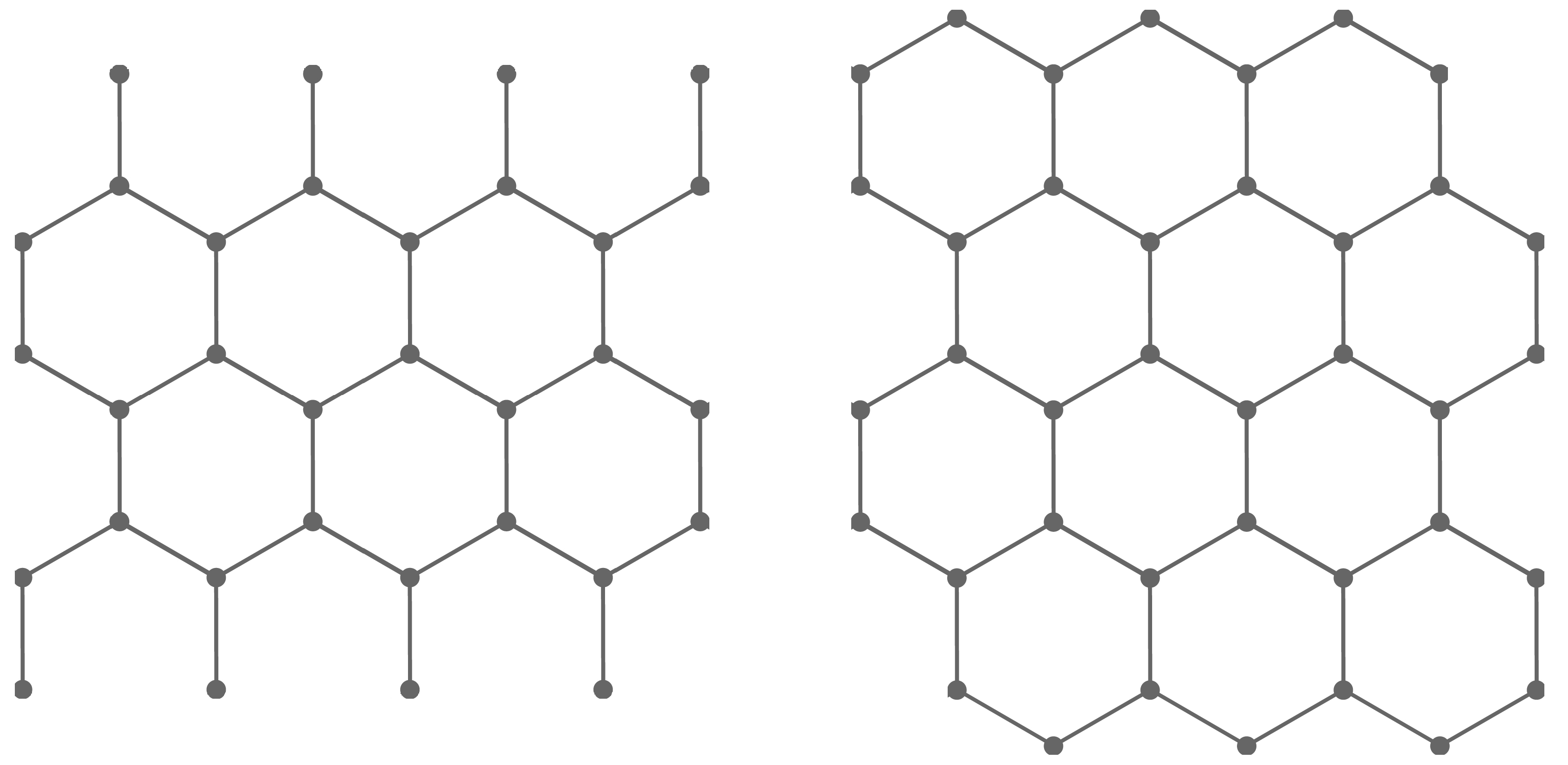}
  \caption{Examples of two finite graphene sheets, with dimensions in
    terms of primitive cells $\left(N_{x},N_{y}\right) =
    \left(4,4\right)$. Along the vertical axis of both sheets are
    armchair edges. The horizontal boundaries are formed by bearded
    edges (left) and zigzag edges (right).  }
  \label{fig:finite_lattices_example}
\end{figure}

We now move to working on graphene sheets, that is, removing the periodic boundary conditions along both axis. A detailed account of a specific version of this set-up together with experimental results has been presented in \cite{Nice} for graphene sheets consisting of armchair boundaries only. Note that such a configuration can not be achieved on rectangular sheets; the simplest configuration has the form of a parallelogram, see \cite{Nice} for further details.
The advantage of such a configuration is that the form of the boundary does not admit so-called `edge-states'. In the following, we will look at the influence of these edge-states in more detail by considering rectangular graphene sheets, such as in Figure \ref{fig:finite_lattices_example}. In addition to armchair edges, here along the vertical axis, these sheets have boundaries formed by bearded edges (left) and zigzag edges (right) along the horizontal axis. These boundaries support edge states, i.e.\ states localised along these edges, with an eigenenergy close to the Dirac energy \cite{Wakabayashi, Yao}. In the following, we will investigate, how the existence of these edge states influences the search.

As in the case of the finite armchair nanotube, the
imposition of Dirichlet boundary conditions at an edge generates
sinusoidal eigenstates. As a result, there are no extended (bulk) eigenstates at the Dirac energy due to the inability to
equate the quantised momenta with the necessary points in
$\underline{k}$-space. We thus need to find other extended eigenstates in the sea of edge states
near the Dirac point to undertake a search in this set-up.

\begin{figure}[t]
  \centering
  \includegraphics[width = 1.0\linewidth]{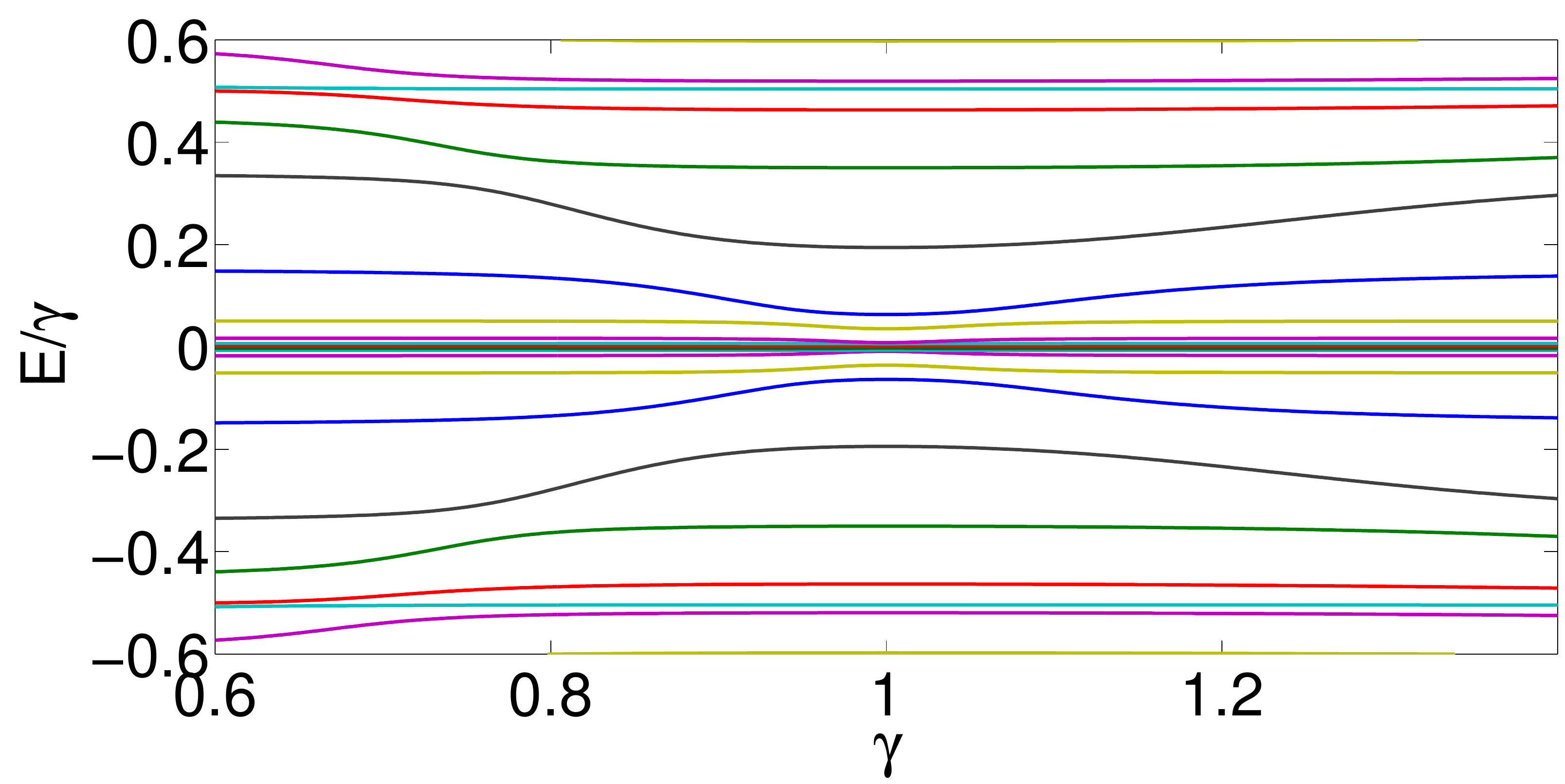}
  \caption{(Color online) Spectrum of triple-bond perturbation search Hamiltonian in
    Equation~\ref{eq:full_search_Hamiltonian} as a function of $\gamma$ for a
    $10\times 10$ cell bearded graphene sheet (N = 200). We focus only
    on the relevant section of the spectrum.}
  \label{fig:spectrum_bearded_3p}
\end{figure}

We mark sites using the triple-bond perturbation such as in
Eq.~(\ref{eq:full_search_Hamiltonian}). Throughout this section we
choose the dimensions in terms of primitive cells of the graphene
sheets, $\left(N_{x} , N_{y}\right) = \left(10, 10\right)$; a bearded
edge sheet thus consists of $N = 200$ sites and a zigzag sheet is
formed of $N = 218$ sites.  The spectra of the search Hamiltonians for
the bearded lattice is shown in Figure~\ref{fig:spectrum_bearded_3p}
constrained to the energy region of interest. The spectrum for zigzag
edges is very similar and is not show here.  One can clearly see an
avoided crossing around the Dirac energy at $\gamma = 1$, the critical
value for this type of perturbation.  There are several states very
close to the Dirac energy; these are the edge states which are all
non-degenerate. Therefore, it is not possible for us to construct a
superposition of degenerate eigenstates which is optimal for
searching.  Rather, our initial starting state must be a single
eigenstates of the unperturbed Hamiltonian $\mathbf{H}_{o} =
-\mathbf{A}$.

Using the search Hamiltonian from
Eq.~(\ref{eq:full_search_Hamiltonian}) and fixing $\gamma = 1$, we
proceed by allowing the system to evolve after being prepared in one
of the unperturbed eigenstates. One finds for both types of lattice,
that the edge states near the Dirac energy fail to produce
localisation behaviour; the probability at the neighboring vertices
of the marked site do not rise above noise levels. Rather, search
behaviour only begins to emerge when we use the first delocalised,
non-edge state as our initial state.  We also note that we
\emph{cannot} search for sites near the zigzag or bearded edges, where
the edge states exist.  Only as we move further into the bulk of the
lattice or along an armchair-type edge, the localisation behaviour
returns.
\begin{figure}[t]
  \centering
  \includegraphics[width =
  1.0\linewidth]{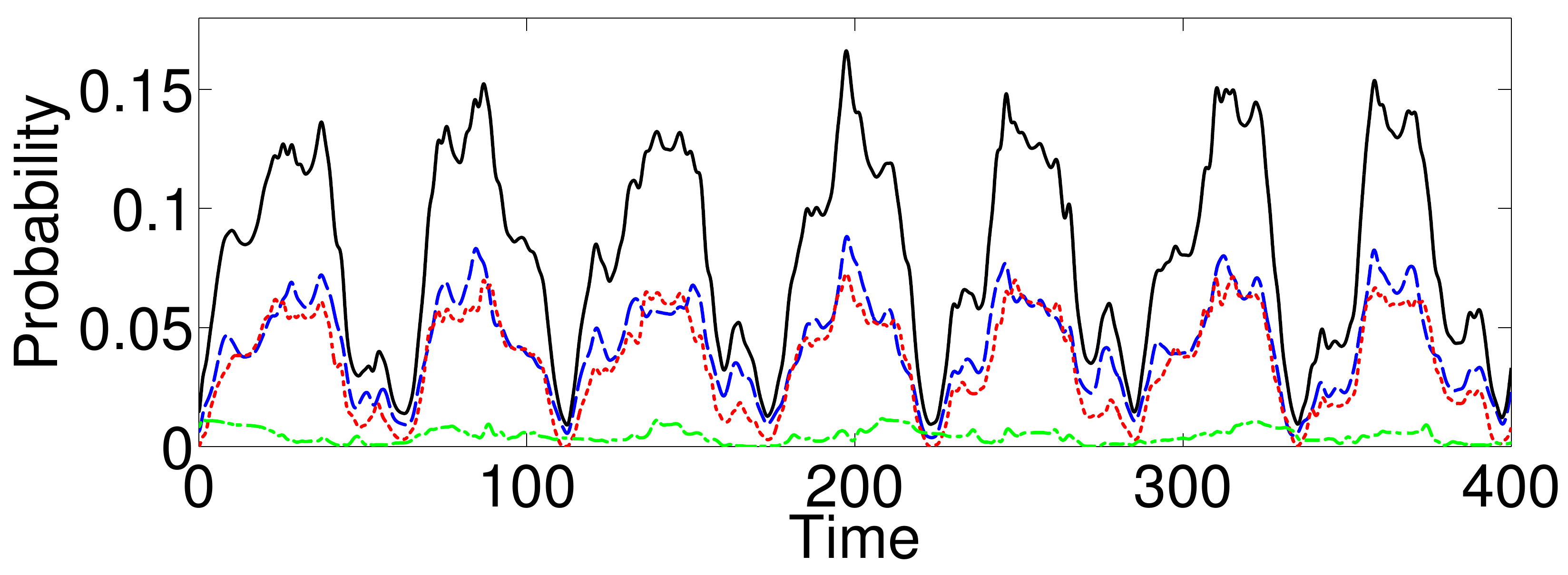}
  \includegraphics[width =
  1.0\linewidth]{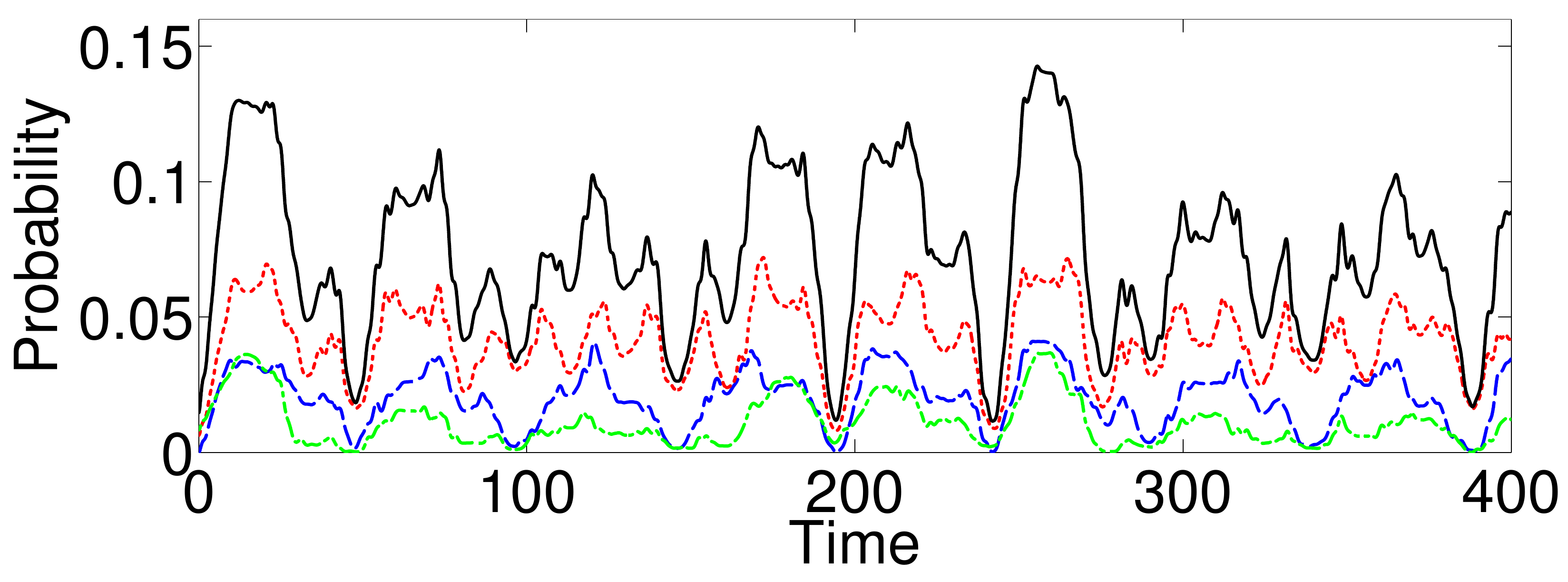}
  \includegraphics[width =
  1.0\linewidth]{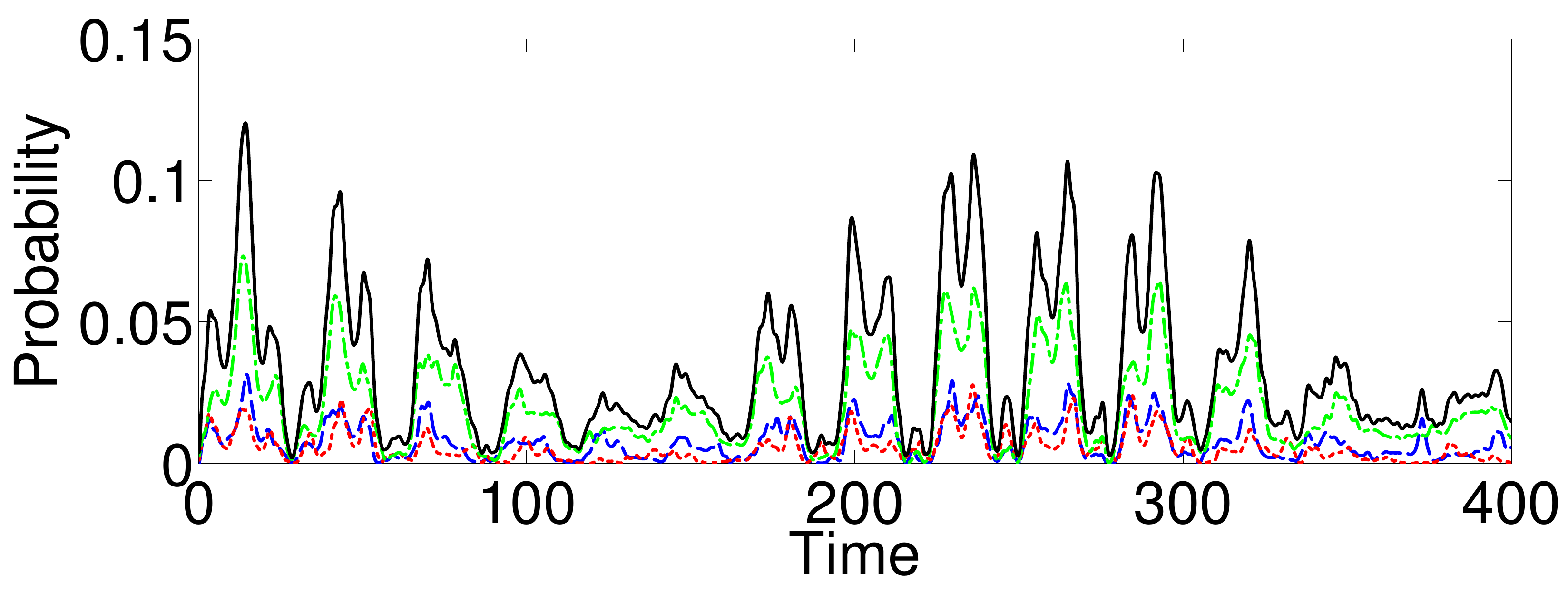}
  \caption{(Color online) Searching on a bearded edge graphene sheet with dimensions
    $\left(N_{x},N_{y}\right) = \left(10, 10\right)$ using the
    Hamiltonian, Eq.~(\ref{eq:full_search_Hamiltonian}). All the
    searches are initialised in the first unperturbed, non-edge
    eigenstate above the Dirac energy.  The locations of the marked
    vertices are: upper - halfway along the left armchair edge, middle
    - centre of the sheet, lower - at the mid-point of the lower
    bearded edge and a third of the way along the vertical axis. (Colors and linestyles as in Figure 
    \ref{fig:armchair_search}) }
  \label{fig:bearded_searches}
\end{figure}
Note that the edge states can be viewed as a kind-of one-dimensional
system and, as we saw from the scaling argument towards the end of
Sec.~\ref{sec:quantumwalksandsearching}, one-dimensional systems imply
an energy spacing between successive energy levels of $E_{n+1}- E_{n}
= \mathcal{O}\left(N^{-1}\right)$. Thus, the perturber state interacts
with many states in a dense part of the spectrum and the search fails.

In Figure~\ref{fig:bearded_searches} we show the search behaviour
arising from marked sites placed at different positions on a bearded
graphene lattice (zigzag lattice types display similar behaviour).
The results are similar to those found for the nanotube searches.
There is a slight increase in variation of signal pattern with
position, and an increase in success probability maxima as we move
towards the centre of the lattice from either of the bearded edges.

\section{Discussion}
\label{sec:discussion}

We have shown that continuous-time quantum search can be done
effectively on a two-dimensional graphene lattice without the use of
internal degrees of freedom. This is achieved by making use of the
conical (linear) dispersion relation in the graphene spectrum. The
search succeeds in time $ T = \mathcal{O}\left(\sqrt{N\ln N}\right)$
with probability $\mathcal{O}\left(1/\ln N\right)$. This is the same
time complexity found in \cite{Aharonov} for discrete-time searches
and in \cite{CG04a} for continuous-time searches. To boost the
probability to $\mathcal{O}\left(1\right)$, $\mathcal{O}\left(\ln
  N\right)$ repetitions are required giving a total time $T =
\mathcal{O}\left(\sqrt{N}\ln^{\frac{3}{2}}N\right)$. Amplification
methods \cite{Gro97a,Tul08,PGKJ09} may be used to reduce the total
search time further.

Our main result focusses on perturbations which involve altering the
hopping potential from a marked site to all three of its
nearest-neighbors equally. We have also demonstrated other types
searches based on perturbing the hopping potential in a cell and the
adding extra sites. We have shown that search mechanisms can be
utilised for the purposes of signal transfer.

Our findings point towards applications in directed signal transfer,
state reconstruction, or sensitive switching. This opens up the
possibility of a completely new type of electronic engineering using
single atoms as building blocks of electronic devices. Our results
demonstrate that a range of nanostructures constructed from graphene
could be used to this end. 

\acknowledgements
We thank Julian B\"{o}hm, Ulrich Kuhl and Fabrice Mortessagne for discussions on how the ideas of \cite{FGT} could
be adapted to more physical systems, which lead to some of the work on additional perturbation types and graphene
flakes.

\appendix
\section{Calculation of
  $Z_{2}\left(S_{\underline{K}/\underline{K}'},1\right)$}
\label{appendix_a}

We give here some details regarding the derivation of
Eq.~(\ref{eq:I2est}) and the logarithmic divergence of $I_{2}$.  It is
clear that the dominant contributions to the $I_{2k}$ summations come
from the vicinity of the Dirac points. Approximating the spectrum
close to the Dirac points, one has
\begin{equation}
  \begin{split}
    I_{2k}=& 2\sqrt{3} N^{k-1} \left[ \sum_{(p,q)\in L}
      \frac{1}{\left(S_{\underline{K}, 11}p^2 + 2S_{\underline{K},
            12}pq+ S_{\underline{K}, 22}q^2 \right)^{k}}
      +\right. \\
    &\left.  \sum_{(p,q)\in L} \frac{1}{\left(S_{\underline{K'},
            11}p^2 + 2S_{\underline{K'}, 12}pq+ S_{\underline{K'},
            22}q^2 \right)^{k}} \right] + O(1) \ .
  \end{split}
\end{equation}
Here the sums over integers $p$ and $q$ is over a rectangular region
$L$ of the lattice $\mathbb{Z}^2$ which is centered at $(0,0)$ and has
side lengths proportional to $\sqrt{N}$ -- the center $(0,0)$,
corresponding to the relevant Dirac point, is omitted from the sum.
As stated in the main text, for $k>1$ the corresponding sums converge which proves Eq.~(\ref{eq:Ikest}).\\
For $k=1$ we will establish constant $C_1$ and $C_2$ such that
\begin{equation}
  C_1 \ln N <  \sum_{(p,q)\in L}
  \frac{1}{S_{\underline{K}, 11}p^2 + 2S_{\underline{K}, 12}pq+
    S_{\underline{K}, 22}q^2 } < C_2 \ln N
  \label{logineq}
\end{equation}  
which then directly leads to Eq.~(\ref{eq:I2est}).  To establish $C_1$
note that because each term in the sum (\ref{logineq}) is positive its
value decreases by restricting it to a square region $- a_1 \sqrt{N}
\le p \le a_1 \sqrt{N}$, $- a_1 \sqrt{N} \le q \le a_1 \sqrt{N}$ which
is completely contained in $L$.  Up to an error of order one the sum
over a square region can in turn be written as a sum over eight terms
of the form
\begin{equation}
  \sum_{p=1}^{a_1 \sqrt{N}} \sum_{q=1}^p 
  \frac{1}{S_{\underline{K}, 11}p^2 \pm 
    2S_{\underline{K}, 12}pq+
    S_{\underline{K}, 22}q^2} \ . 
\end{equation}
For fixed $p$ we can find $q_{\mathrm{max}}$ such that
\begin{multline}
  \sum_{q=1}^p \frac{1}{S_{\underline{K}, 11}p^2 \pm
    2S_{\underline{K}, 12}pq+
    S_{\underline{K}, 22}q^2}>\\
  \frac{p}{S_{\underline{K}, 11}p^2 \pm 2S_{\underline{K},
      12}pq_{\mathrm{max}}+ S_{\underline{K}, 22}q_{\mathrm{max}}^2}\
  .
\end{multline}
We may choose $q_{\mathrm{max}}= b_1 p$ for some constant $b_1\ge 0$,
so
\begin{equation}
  \sum_{p=1}^{a_1 \sqrt{N}} \sum_{q=1}^p \frac{1}{S_{\underline{K}, 11}p^2 \pm 
    2S_{\underline{K}, 12}pq+
    S_{\underline{K}, 22}q^2} >
  c \sum_{p=1}^{\sqrt{N}} \frac{1}{p}
\end{equation}  
which diverges as $\ln N$.

Establishing $C_2$ and the corresponding logarithmic bound from above,
and thus proving Eq.~(\ref{eq:I2est}), follows the same line by first
extending the sum to a square of side length $2 a_2 \sqrt{N}$ that
completely contains $L$ and then establishing
\begin{multline}
  \sum_{q=1}^p \frac{1}{S_{\underline{K}, 11}p^2 \pm
    2S_{\underline{K}, 12}pq+
    S_{\underline{K}, 22}q^2}<\\
  \frac{p}{S_{\underline{K}, 11}p^2 \pm 2S_{\underline{K},
      12}pq_{\mathrm{min}}+ S_{\underline{K}, 22}q_{\mathrm{min}}^2}\
\end{multline}
with $q_{min}=b_2 p$.

\end{document}